\documentclass{report}

\usepackage{arxiv}

\usepackage[utf8]{inputenc} % allow utf-8 input
\usepackage[T1]{fontenc}    % use 8-bit T1 fonts
\usepackage{hyperref}       % hyperlinks
\usepackage{url}            % simple URL typesetting
\usepackage{booktabs}       % professional-quality tables
\usepackage{amsfonts}       % blackboard math symbols
\usepackage{nicefrac}       % compact symbols for 1/2, etc.
\usepackage{microtype}      % microtypography
\usepackage{lipsum}

\usepackage{lineno}
\usepackage{amsmath}

\usepackage{tabularx}
\usepackage[ruled]{algorithm2e} % For algorithms
\usepackage{algorithmic}

\SetAlFnt{\small}
\SetAlCapFnt{\small}
\SetAlCapNameFnt{\small}
\SetAlCapHSkip{0pt}
\IncMargin{-\parindent}

\usepackage{mathtools}
\usepackage{tikz}
\usetikzlibrary{calc}

\usepackage{epsfig}
\usepackage{epstopdf}
\usepackage{textcomp}
\usepackage{xcolor}
\usepackage{caption}
\usepackage{subcaption}
\usepackage[nottoc,notlof,notlot]{tocbibind}

\title{Investigating Image Applications Based on Spatial-Frequency Transform and Deep Learning Techniques}

\author{
  Qinkai Zheng\\
  Telecom Paris, Institut Polytechnique de Paris\\
  Palaiseau, 91120, France\\
  \texttt{qinkai.zheng@telecom-paris.fr} \\
  [1em] Supervisors: \\
  Han Qiu, Gerard Memmi, and Isabelle Bloch, \\
  Telecom Paris, Institut Polytechnique de Paris\\
  Palaiseau, 91120, France\\
  \texttt{\{han.qiu, gerard.memmi, isabelle.bloch\}@telecom-paris.fr}
}

\begin{document}
\maketitle

\begin{abstract}
This is the report for the PRIM project in Telecom Paris. 
This report is about applications based on spatial-frequency transform and deep learning techniques.
In this report, there are two main works.
The first work is about the enhanced JPEG compression method based on deep learning. 
we propose a novel method to highly enhance the JPEG compression by transmitting fewer image data at the sender's end. 
At the receiver's end, we propose a DC recovery algorithm together with the deep residual learning framework to recover images with high quality.
The second work is about adversarial examples defenses based on signal processing. 
We propose the wavelet extension method to extend image data features, which makes it more difficult to generate adversarial examples.
We further adopt wavelet denoising to reduce the influence of the adversarial perturbations.
With intensive experiments, we demonstrate that both works are effective in their application scenarios.

% In the field of image processing, the transformations in the frequency domain are widely used in various aspects, like image analysis, image compression, image restoration, etc. 
% By extracting the rate of change of pixels, transformations like Fourier transform and wavelet transform can provide more information other than the spatial one. 
% Recently, deep learning techniques become popular in image processing. 
% The end-to-end deep learning framework on the spatial domain achieves surprising performance compared with traditional methods. 
% In this report, we investigate the possibilities of combining image processing in the frequency domain with deep learning techniques. 
% On one hand, deep learning techniques can be used to enhance existing methods that use transformations in the frequency domain. 
% We propose a novel method to enhance the JPEG compression by using the DC recovery algorithm together with the deep residual learning framework. 
% On the other hand, transformations like wavelet transform can be used to enhance the robustness of deep learning models. 
% We propose a novel defense against adversarial examples by using the wavelet extension method. 
% With intensive experiments, we demonstrate that both methods are effective in their application scenarios.

\end{abstract}

\tableofcontents{}

\chapter{Introduction}

In the field of image processing, an image is often represented in the form of a 2D matrix where each element of the matrix stands for the intensity of a certain pixel. 
The intensity distribution of the entire image represents the image in the spatial domain.
It is intuitive to manipulate the pixel value in the spatial domain to process the image.
Other than the spatial domain, one can also apply transformations in the frequency domain.
The frequency for an image represents the rate of change of the image signal.
Transformations like discrete cosine transform and wavelet transform calculate the quantity of signal that lies in the given frequency bands. 
This kind of transformations in the frequency domain provides more information than the spatial domain, which makes it easier to detect more important features. 
They are widely used in various aspects in the field of image processing and achieve significant results.
For example, wavelet features are used for texture analysis in image retrieval~\cite{manjunath1996texture}.
In image compression, new compression methods are proposed based on the compression of wavelet coefficients~\cite{devore1992image}.

Recently, a big number of deep learning techniques are introduced in image processing.
Since the appearance of AlexNet~\cite{krizhevsky2012imagenet}, the convolutional neural networks (CNNs) have shown the superiority over traditional methods in different image processing tasks including image classification~\cite{krizhevsky2012imagenet}, image denoising~\cite{zhang2017beyond}, image compression~\cite{balle2016end}, etc.
CNNs apply 2D convolution on the image with a small kernel to extract local features between pixels. 
CNNs have a large learning capacity that can be controlled by the depth and the kernel size. 
In~\cite{zhang2017beyond}, CNNs are used together with residual learning framework to denoise images with high effectiveness.
In~\cite{balle2016end}, two CNNs are seamlessly integrated into an end-to-end image compression framework. 
These approaches achieved surprising results with the help of the architecture of CNNs and abundant computational power.
Noticed that these approaches always operate directly in the spatial domain in an end-to-end manner, the possibility of using transformations in the frequency domain is not well investigated.

In this report, we investigate the possibilities of combining transformations in the frequency domain with deep learning techniques in two different applications.
On one hand, deep learning techniques can be used to enhance the existing method that is based on transformations in the frequency domain, e.g. JPEG compression based on discrete cosine transform. 
We propose a novel method to enhance the JPEG compression by using the DC coefficient recovery algorithm together with the deep residual learning framework. 
This method can recover images in the receiver's end by only using 4 DC coefficients and all AC coefficients of original images, which results in a 40\% reduction of transmission data.
On the other hand, transformations like wavelet transform can be used to enhance the robustness of deep learning models. 
We propose a novel defense against adversarial examples by using the wavelet extension method.
This method extracts image structures and basic elements by using wavelet transform and extend the dimension of input images.
By preventing adversaries from generating adversarial examples that are effective on both parts, the method increases the robustness of target deep learning models.
By doing intensive experiments, we demonstrate that both methods are effective in their application scenarios.

The roadmap of this report is as follows: in Chapter~\ref{chap:jpeg}, we propose an enhanced JPEG compression method based on accurate DC recovery and deep residual learning; in Chapter~\ref{chap:wave}, we introduce a defense mechanism for deep learning models against adversarial examples by using the wavelet extension method; in Chapter~\ref{chap:discuss}, we further discuss the limits of these methods and the potential of improvement.

\chapter{Deep Residual Learning-based Enhanced JPEG Compression Method}
\label{chap:jpeg}

With the development of big data and network transmission technology, multimedia contents such as images or videos are widely transmitted in all kinds of networks. 
However, traditional multimedia compression methods based on spatial-frequency transformation and coding techniques are not correspondingly developed and touching the limits of compression ratio which brings the transmission latency issues for multimedia data. 
In this chapter, we propose a novel method to highly enhance the JPEG compression by transmitting only four DC coefficients and all AC coefficients of one image at the sender's end.  
At the receiver's end, firstly, we propose a state of the art DC recovery method to preprocess the received four DC coefficients and all AC coefficients. 
Then, a deep residual learning model is built to remove the block artifacts for the preprocess results which can provide a high level of image quality.
By performing these two steps, our proposed method could recover the transmitted only four DC coefficients and all AC coefficients with more than 31 dB considering PSNR while transmitting only 60\% data of the original JPEG images. 

\section{Introduction}
\label{intro}

During the last decades, the digital data generation and transmission have been evolving and increasing significantly~\cite{qiu2019user}. 
Multimedia data files, such as images or videos, are rapidly produced and transmitted through all kinds of networks. 
As pointed in~\cite{miao2016performance}, multimedia networks are widely deployed and transmission latency is becoming serious issues for multimedia data sharing. 
However, although the multimedia data volume increased rapidly, the corresponding compression algorithms are not well developed in the meantime.

Most of today's multimedia data are still compressed based on information theory. 
\textit{Discrete Cosine Transform (DCT)}~\cite{ahmed1974discrete} is the most widely used transform in multimedia coding, which covers image and video coding standards such as JPEG, MPEG-1/2, MPEG-4, AVC/H.264~\cite{schwarz2007overview}, and the more recent HEVC~\cite{zeng2013tutorial}. 
DCT is a Fourier-like transform, which was first proposed by~\cite{ahmed1974discrete} to compact most energy of a highly-correlated discrete signal into a few coefficients
The purpose of DCT is to perform de-correlation of the input signal and to present the output in the frequency domain just like other transformation algorithms. 
The DCT could well represent a signal with only the cosine-series expansion which gets rid of the complex numbers in the Fourier transform. 
Deploying DCT in the multimedia data compression is basically to determine the importance levels of the DCT coefficients considering the visual effects and then to use the quantization step to compress these coefficients with unrecoverable loss. 

Most multimedia compression standards working in frequency domain apply the DCT transform to smaller blocks sequentially to reduce the overall time complexity of the transformation. 
For instance, the JPEG standard~\cite{wallace1992jpeg} deploys DCT algorithms on two-dimension (DCT-2D) at a small block level with the block size of $8 \times 8$ pixels.
This DCT-2D for one $8 \times 8$ block will generate 64 coefficients with each coefficient carries distinct information of the transformed signal. 
Among all DCT coefficients, the first (i.e., the one with the lowest frequency) coefficient is called the DC coefficient while the rest are called AC coefficients. 
DC coefficient is considered the special one because it carries the average intensity of the transformed signal which is the average value of the pixel values of the 64 pixels in this $8 \times 8$ block. 
The compression algorithm will then compress the DC and AC coefficients with quantization steps to ignore many of the AC coefficients. 

One ground truth of the multimedia compression is that compressing more AC coefficients will achieve a higher compression ratio but lead to worse image quality. 
JPEG standard also has different quantization tables for changing how many AC coefficients will be kept according to compression requirements. 
However, on the other hand, compression on the DC coefficients could also be a promising approach to achieve a higher compression ratio. 
For instance, the authors in~\cite{chen2017dc} showed a video compression method based on the prediction of the DC coefficients. 
Also,~\cite{qiu2019dc} showed DC coefficients of JPEG images can be guessed with only AC coefficients. 

These research approaches inspired that the JPEG image could be compressed with keeping only AC coefficients at the sender's end if the DC coefficients can be then recovered at the receiver's end. 
However, the main two approaches introduced in Section~\ref{background} for recovering DC coefficients are all suffering from the prediction errors that cannot be overcome~\cite{qiu2019dc, uehara2006recovering}. 
The prediction errors are due to the basic observed theory of the pixel passing in~\cite{uehara2006recovering} is not always true (see Section~\ref{problemDef1}). 

Therefore, our main contributions are: (1) we propose a state of the art DC coefficient recovery method for JPEG based on only four DC coefficients and all AC coefficients; (2) we propose the \textit {Deep Learning (DL)} model to solve the block artifacts that cannot be overcome by the traditional DC recovery methods; (3) by combining the traditional DC recovery and the DL model, we implement an enhanced JPEG compression method by transmitting only 60\% of the JPEG image with PSNR achieving more than 30 dB.

The roadmap for this chapter is as follows: 
in Section~\ref{background}, some related research background is presented; 
in Section~\ref{systemDes}, we illustrate how to use the proposed method; 
in Section~\ref{problemDef}, we present the preprocess DC recovery method;
in Section~\ref{DeepLea}, we present the deep residual learning model; 
in Section~\ref{Experiment}, we explain the experiment details and evaluate the method; 
in Section~\ref{discuss1}, we discuss about the research motivation; 
in Section~\ref{conclusion}, we conclude our work.

\section{Research background}
\label{background}

In this section, we list the research background mainly on two aspects. 
Firstly, the DC coefficients are explained with a more practical definition. 
We illustrate the initial research motivation of why DC coefficients are special and why the research of recovering DC coefficients from the rest AC coefficients is performed. 
Then, we listed the two main approaches of the previous work for recovering the DC coefficients which are based on the block by block propagation methods and based on image spatial domain optimization respectively. 

\subsection{Practical definition of the DC coefficients}

In this subsection, we will briefly explain the very practical definition of the DC coefficients in the DCT. 
Since DCT has different types shown in~\cite{kresch1999fast}, the most popular DCT algorithm is a two-dimensional symmetric variation of the transform that operates on $8 \times 8$ blocks (DCT $8 \times 8$) and its inverse (iDCT $8 \times 8$). This DCT $8 \times 8$ is utilized in JPEG compression routines~\cite{wallace1992jpeg} and has become an important standard in image and video compression steps.

\begin{figure}[!htbp]
\centering
\includegraphics[width=0.8\textwidth]{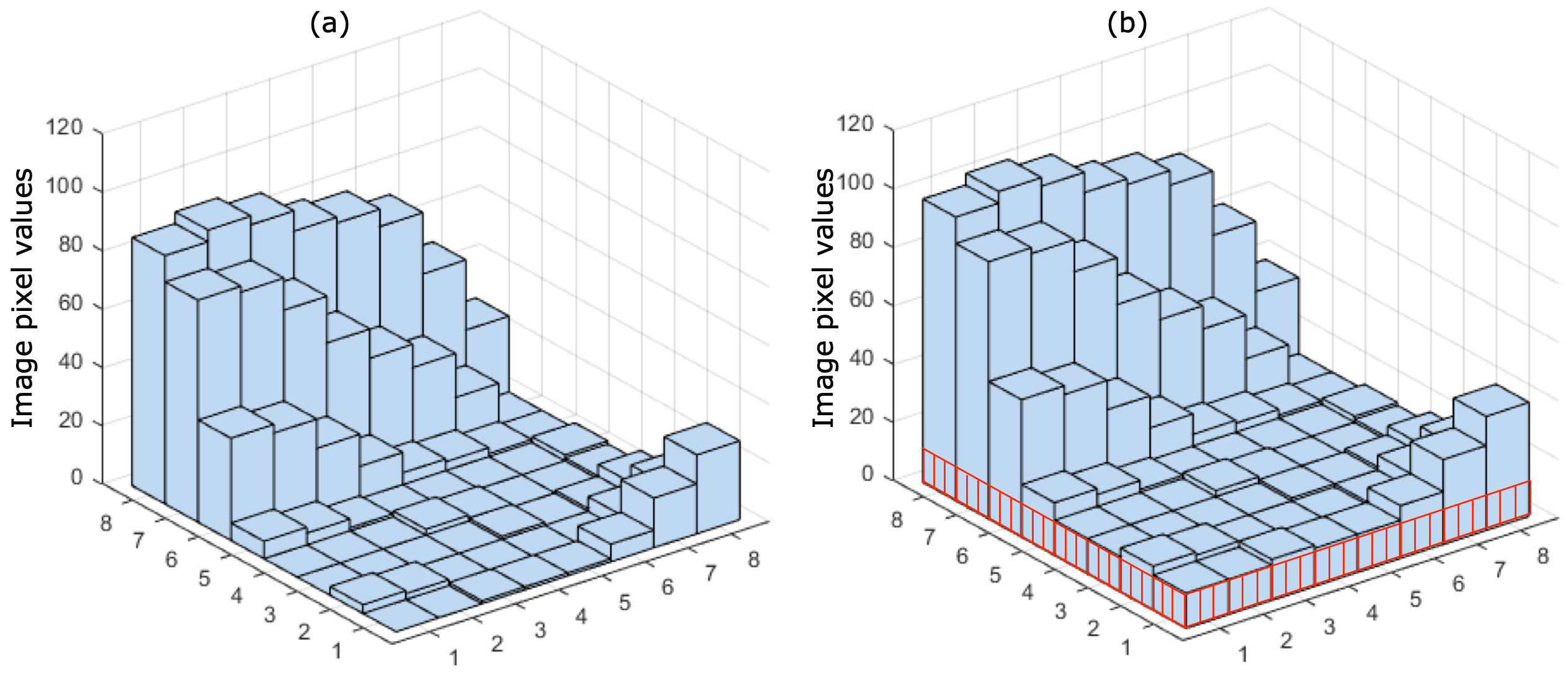}
\caption{An example of the definition of DC coefficient in a $8 \times 8$ block: (a) original distribution of pixel values (b) distribution of pixel values when DC coefficient increased: all pixel values increased with the same value}
\label{pixeldist1}
\end{figure}

According to the definition of the DCT transformation~\cite{wallace1992jpeg}, the DC coefficient is the average value of the input elements. 
Thus, the DC coefficients of the DCT transform of image blocks represent the mean values of the pixel values in the corresponding image blocks. 
We illustrate this definition by an example of one $8 \times 8$ block shown in~\figurename~\ref{pixeldist1}. 
In the~\figurename~\ref{pixeldist1} (a), we list the pixel values on the $z$-axis for the original block. 
Then, we add the DC coefficient in the DCT result and do the iDCT to get the pixel values with only the DC coefficient increased in~\figurename~\ref{pixeldist1} (b). 
The pixel value distribution is not changed in the ~\figurename~\ref{pixeldist1} (a) and (b) while every pixel value is added with the same value. 

%\vspace{-1em}
%\vspace{-1.5em}

Thus, for one JPEG image, if all DC coefficients in all $8 \times 8$ blocks are changed to zero, the relative difference of the pixel values still exist without any loss. 
We could consider that for the pixel value distribution on one JPEG image with all DC coefficients are zeros means every $8 \times 8$ block is still keeping the distribution inside the block but the difference is the average values of the different blocks. 
Based on this practical explanation for the DC coefficients, the motivation of this research is to transmit only the AC coefficients in each block at the sender's end and to recover the corresponding DC coefficients at the receiver's end to realize higher compression ratio for JPEG images.

\subsection{Previous approaches on the DC coefficient recovery}
\label{previous}

The very initial question of the DC coefficients recovery was proposed by~\cite{uehara2006recovering} since the early stage of SE methods used to protect the DC coefficients of each $8\times 8$ block to further protect the image content. 
Such image protection methods are deployed on the JPEG images or other image formats with DCT transformation. 
The implementation of such methods showed a hard visual degradation for the images but the authors in~\cite{uehara2006recovering} showed that the DC coefficients of the $8\times 8$ JPEG blocks can be guessed by the rest AC coefficients. 

The guessing method is based on the observed statistical distribution of the image property in~\cite{uehara2006recovering}: in general, the difference signal at the pixel level and the DC coefficients has been modeled as a zero-mean Laplacian distributed variate. 
The distribution is generally narrow with a small value of variance. 
Therefore, we could assume the value of variance as zero to determine the difference of the DC coefficients of the neighbor $8\times 8$ block. 
This method is experimented in~\cite{uehara2006recovering} for the DC coefficients recovery in JPEG images and further improved in~\cite{qiu2019dc, qiu2019dc2}. 
However, one serious issue existed in this method since there will be tiny errors if we use the DC coefficient in the $8\times 8$ block $B_{(i,j)}$ to guess the neighbor $8\times 8$ block $B_{(i+1,j)}$ and this error will propagate. 
As shown in~\cite{qiu2019dc, qiu2019dc2}, although the recovery results have an acceptable PSNR value and the image contents can be observed, the error propagation will lead to a lot of edges in the image contents. 

The other approach is based on the optimization algorithms such as shown in~\cite{li2011recovering} since this question can be solved by the \textit{Linear Programming (LP)} technique for not only the basic DC recovery problem and the general DCT coefficients recovery problem. 
The authors in~\cite{li2011recovering} also showed the recovery for the image content even more than 15 AC coefficients are also missing. 
In a bitmap case, more than half of the low-frequency coefficients are missing, a rough recovery for the image content is still possible~\cite{qiu2018ssic}.
However, this approach does not suit our case since the image we want to deal with is JPEG images. 
In JPEG images, as there are the quantization and rounding step~\cite{pennebaker1992jpeg}, most of the DCT coefficients are rounded to zeros (could be more than 90\% of the DCT coefficients in the JPEG images are zeros as pointed in~\cite{qiu2019dc}).
Although this LP approach can optimize the calculation and remove the edges due to the error propagation, the recovery results lost much detailed information. 
The summarized problem definition and the main motivation of our research of using DL can be seen in Section~\ref{problemDef}.

\section{System design}
\label{systemDes}

In this section, we show how to use the proposed method in the practical JPEG image transmission scenario. 

\begin{figure}[!htbp]
\centering
\includegraphics[width=0.5\textwidth]{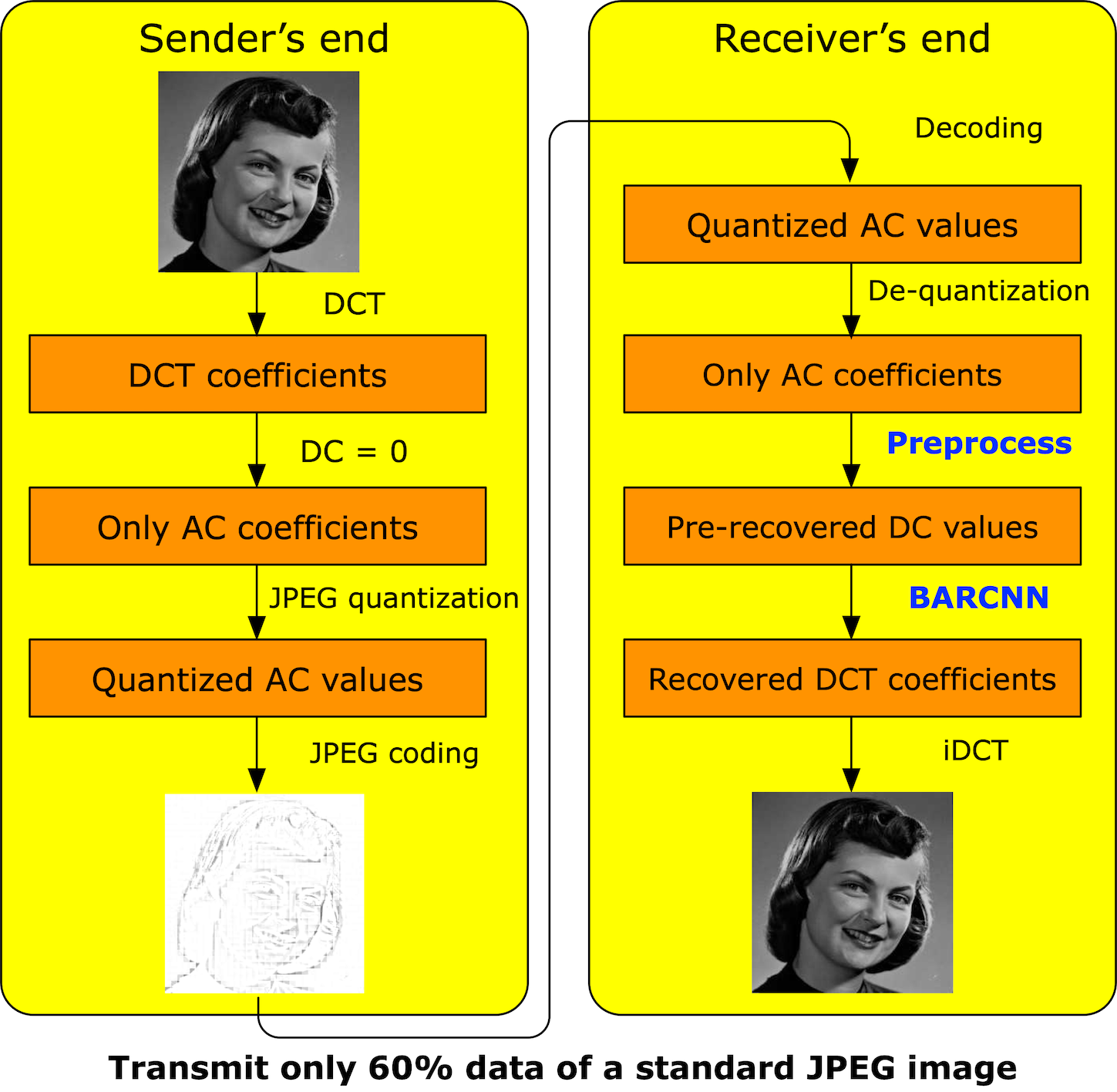}
\caption{An example of how the proposed method can be used with only 60\% data transmitted for one JPEG image.}
\label{system1}
\end{figure}

As shown in~\figurename~\ref{system1}, at the sender's end, the JPEG image can be compressed without the DC coefficients. 
More specifically, for the standard JPEG compression process, we keep the same transformation step (DCT-2D) and the encoding process. 
The only difference is that the DC coefficients in every $8 \times 8$ will be removed and replaced with zeros. 
In our work, we keep only four DC coefficients of the $8 \times 8$ blocks in the four corners of the image as the reference for the recovery step. 
For instance, for an image with a size of $256 \times 256$, there are $32 \times 32$ blocks with each block contains one DC coefficient. 
For such an image, there are initially 1024 DC coefficients in total for the standard JPEG compression but now there suppose to be only 4 DC coefficients to be transmitted since 1020 DC coefficients are changed to zeros. 

At the receiver's end, the JPEG decoding process is the first performed to get the correct AC coefficients and the four DC coefficients of the four corner blocks. 
A preprocessing step will be deployed to have a prediction of the missing DC coefficients. 
This preprocessing step is proposed based on improving the previous work~\cite{qiu2019dc} which can roughly recover the DC coefficients based on the pixel passing theory in~\cite{uehara2006recovering}. 
However, as pointed in~\cite{qiu2019dc}, there will be block artifacts (not noise) in the recovered image due to the error propagation. 
Then, a deep residual learning model based on CNN, which is proposed in this paper, will be used to further recover the DC coefficients precisely by removing the block artifacts and provide better image quality. 
Through this operation, the transmission data for a $256 \times 256$ JPEG image are highly reduced (only 60\% compared with standard JPEG images). 
For different images, the compression performance will be around 60\% for many JPEG datasets (see Section~\ref{compress1}).

\section{Preprocess method design}
\label{problemDef}

In this section, we illustrate the preprocess method in~\figurename~\ref{system1}. 
Firstly, we describe the problems of error propagation existed in the traditional approach based on the pixel passing distribution theory in~\cite{uehara2006recovering}. 
Secondly, we present our improvement to overcome this error propagation by improving the method proposed in~\cite{qiu2019dc} which still suffers from the error propagation situation. 
This question is then summarized into a more precise statistical problem and we propose the motivation of using DL based approach to solve.

\subsection{Problem of traditional approach}
\label{problemDef1}

As discussed in Section~\ref{previous}, there are two main approaches to solving the DC coefficient recovery problems. 
In fact, in this special use case that only DC coefficients are missing for JPEG images, all AC coefficients have remained which means for each $8 \times 8$ block, the relative pixel value distribution is accurately remained. 
If we use the method operated in the spatial domain or try to optimize the image content with the LP approach, even the recovered image is more smooth, most of the AC coefficients are changed. 
Therefore, we not only failed to recover the accurate DC coefficient but also introduce the errors for the AC coefficients. 
Thus, we use the first approach that is focused on how to recover the image content by accurately recovering the DC coefficients without changing the AC coefficients. 

\begin{figure}[!htbp]
\centering
\includegraphics[width=0.8\textwidth]{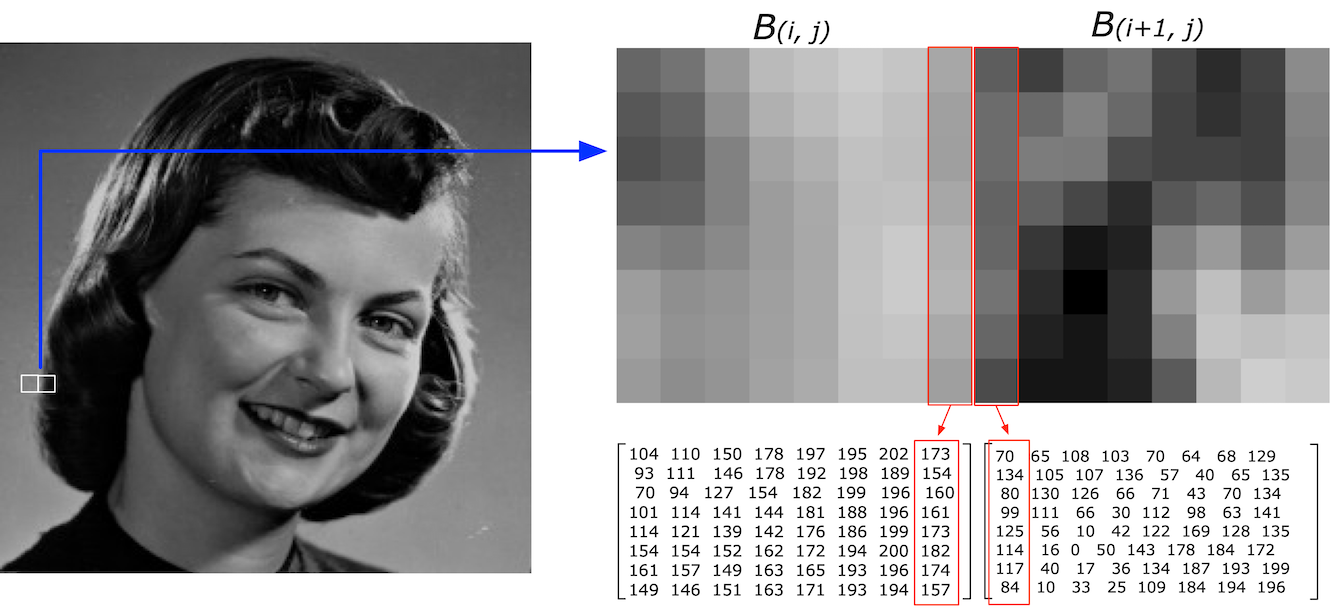}
\caption{An example of the real case of pixel value distribution that does not fit the zero-mean Laplacian distributed variance in neighbour two blocks~\cite{uehara2006recovering}.}
\label{errorExample1}
\end{figure}

We firstly indicate that the basic observed theory in~\cite{uehara2006recovering} does not fit the practical scenario with an example shown in~\figurename~\ref{errorExample1}. 
These two $8 \times 8$ blocks are picked from an image and the neighbor two vectors of the neighbor pixels are very different. 
For this case, the method in~\cite{qiu2019dc} which is always trying to find the DC value to achieve the minimum \textit{Mean Square Error (MSE)} of two neighbor blocks will introduce wrong predictions for the DC value since the real case is that the MSE of the real case is very large. 

In~\figurename~\ref{errorExample1}, if we know the DC value of the block $B_{(i,j)}$ is 80, based on the min MSE method in~\cite{qiu2019dc}, the predicted DC value of the block $B_{(i+1,j)}$ is 82. 
The difference of the two neighbour vectors in~\figurename~\ref{errorExample1} is $[7,-13, -15, 3, 9, -1, 2, 19]$ while the real DC value of the block $B_{(i+1,j)}$ is 49 and the difference of the neighbour 8 pixels is $[103, 20, 80, 62, 48, 68, 57, 73]$. 
The fact is that the pixel passing of these two neighbor blocks is not the zero-mean Laplacian distributed variate. 
On the opposite, the variance between these two blocks is very intensive. 
This is the reason why the image recovery methods based on DC recovery in~\cite{uehara2006recovering,qiu2019dc} will have the propagated errors and the block artifacts in the image. 
Thus, the problem to be solved is that for cases shown in~\figurename~\ref{errorExample1} which does not fit the observed statistical theory in~\cite{uehara2006recovering}.

\subsection{Improved DC recovery method}

In this subsection, we propose the preprocess step in~\figurename~\ref{system1} based on improving the previous approach in~\cite{qiu2019dc}. 
According to our observation, the ratio that the pixel passing of the neighbor $8\times 8$ blocks fit the zero-mean Laplacian distributed variate theory in~\cite{uehara2006recovering} is more than 90\%. 
Such that we first tried to improve the method in~\cite{qiu2019dc} by calculating one $8\times 8$ block's pixel passing for multiple directions. 
For instance, for one block $B_{(i,j)}$, we first try to predict its DC coefficient by calculating the smoothest pixel passing compared with its left neighbor block $B_{(i-1,j)}$. 
Then, we predict its DC coefficient by its upper neighbor block $B_{(i,j-1)}$. 
The average value of the two predicted DC coefficients will be seen as the predicted DC coefficient for block $B_{(i,j)}$. 
The reason for such performing is that for the blocks that do not have smooth pixel passing with its left neighbor blocks, they still have a 90\% chance to have a smooth pixel passing with its upper neighbor blocks. 
Such that the average calculation will make up the errors of the DC prediction in one direction as pointed in~\figurename~\ref{errorExample1}. 

\begin{figure}[!htbp]
\centering
\includegraphics[width=0.6\textwidth]{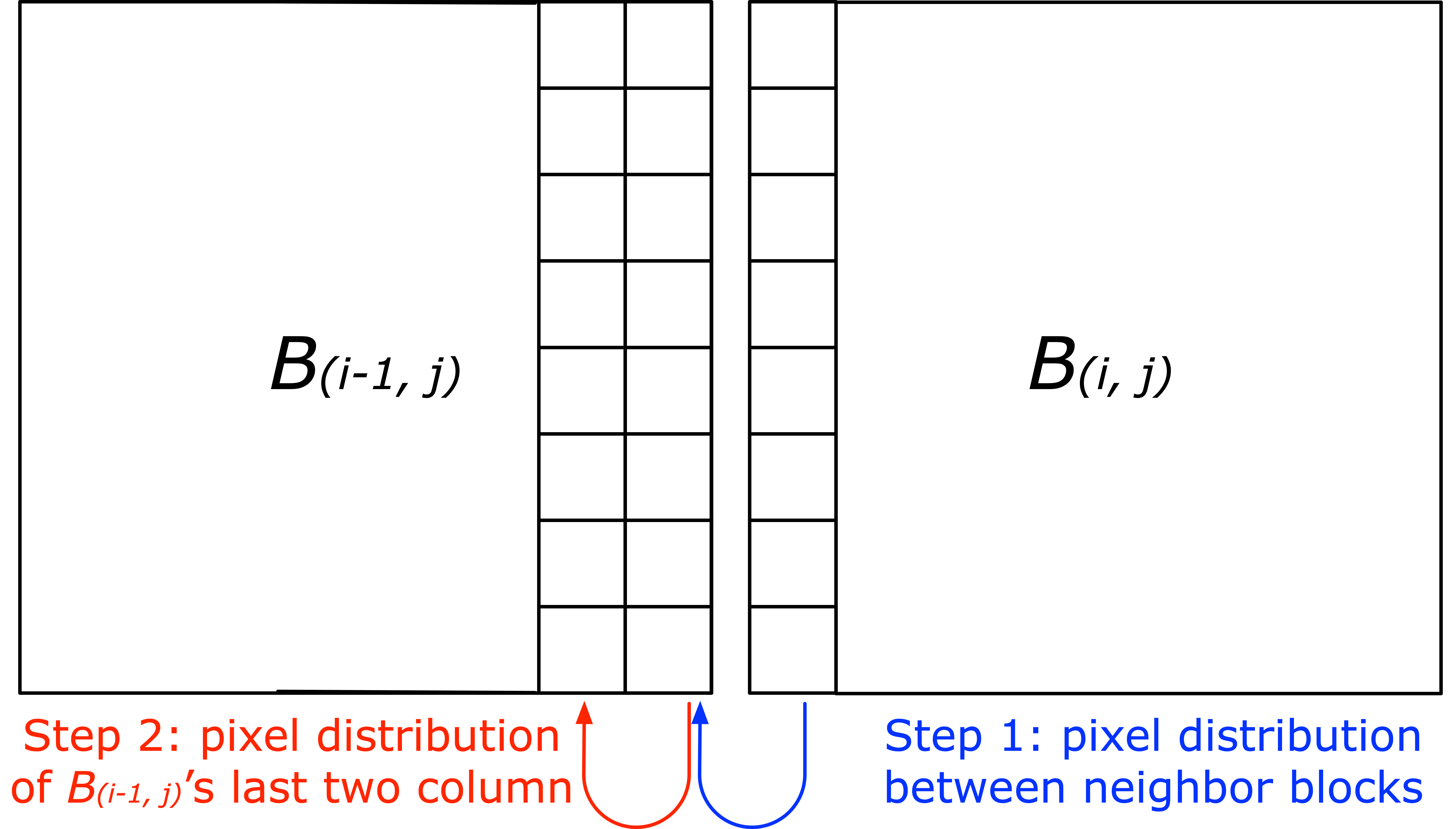}
\caption{An example of the improved DC prediction method proposed in our work: step 1: predict DC coefficient of $B_{(i,j)}$ based on the pixel passing with neighbor block $B_{(i-1,j)}$; step 2: predict DC coefficient of $B_{(i,j)}$ based on the pixel passing trends of the last two columns of neighbor block $B_{(i-1,j)}$.}
\label{twosteps1}
\end{figure}

We also tried to fix this situation by also calculating one DC coefficient based on the pixel passing of the last two columns of its left neighbor block. 
The method as shown in~\figurename~\ref{twosteps1} is to find one DC coefficient $D(i,j)$ that can make the pixel passing distribution of the blocks $B_{(i,j)}$ and $B_{(i-1,j)}$ most similar with the pixel passing distribution of last two column pixels in $B_{(i-1,j)}$.

In our experimentation, we deployed the scheme in~\figurename~\ref{twosteps1} for two directions (compute for $B_{(i,j)}$ with $B_{(i-1,j)}$ and $B_{(i,j-1)}$) as shown in~\tablename~\ref{tab:notations} and Algorithm~\ref{algo1}.

\begin{table}[!htbp]
\centering
\caption{Major notations used in the algorithm and their definitions.}
\label{tab:notations}
\begin{tabular}{l l }
\hline
Notation                  & Definition  \\ \hline
$B_{(i,j)}$              & $8\times 8$ block with location $(i, j)$ \\
$C_{(i,j)}$              & 64 DCT coefficients of block $B_{(i,j)}$ \\ 
$A_{(i,j)}$              & 63 AC coefficients of block $B_{(i,j)}$ \\ 
$D_{(i,j)}$              & DC coefficient of block $B_{(i,j)}$ \\
$P_{(i,j)}$              & Pixel values of block $B_{(i,j)}$ \\
$\tilde{D}_{(i,j)}$      & Estimated DC coefficient of block $B_{(i,j)}$ \\
$Q_{50}$              & Quantization table of JPEG used in our work \\ 
$v_n$                 & Number of $8\times 8$ blocks in vertical direction \\
$h_n$                 & Number of $8\times 8$ blocks in horizontal direction \\
$iDCT$                & Inverse Discrete Cosine Transform \\ 
$Concat$              & Concatenation of two arrays \\
$\mathcal{MSE}$       & MSE between two neighbour blocks defined in~\cite{qiu2019dc} \\

\hline
\end{tabular}
\end{table}

% \begin{minipage}{.7\linewidth}
\label{algo1}
\begin{algorithm}
\caption{DC coefficients prediction from upper-left to bottom-right corner.}
 {\bf Input:} AC coefficients $A_{(i,j)}$ with $(i,j)\in \{v_n, h_n\}$, DC coefficient $D_{(1, 1)}$ of upper-left block\\
 {\bf Output:} Recovered DC coefficients $\tilde{D}_{(i,j)}$ with $(i,j)\in \{v_n, h_n\}$ \\

\begin{algorithmic}
\STATE $\tilde{D}_{(1,1)}=D_{(1,1)}$
\FOR{$i \gets 1$ to $v_n$}
    \FOR{$j \gets 1$ to $h_n$}
        \STATE /*~Find the optimal DC that minimize loss.~*/
        \FOR{$DC \gets -64$ to 64} 
            \STATE /*~Calculate spatial domain value.~*/
            \STATE $P(i,j)=iDCT(Concat(DC, A_{(i,j)})*Q_{50})$ 
            \STATE $P(i-1,j)=iDCT(Concat(\tilde{D}_{(i-1,j)}, A_{(i-1,j)})*Q_{50})$
            \STATE $P(i,j-1)=iDCT(Concat(\tilde{D}_{(i,j-1)}, A_{(i,j-1)})*Q_{50})$
            \STATE $\mathcal{MSE}(DC)=\mathcal{MSE}(P_{(i,j)}, P_{(i-1,j)}) + \mathcal{MSE}(P_{(i,j)}, P_{(i,j-1)})$
        \ENDFOR
        \STATE $\tilde{D}_{(i,j)}=argmin_{DC} \{\mathcal{MSE}(DC)\}$
    \ENDFOR
\ENDFOR
\end{algorithmic}
\end{algorithm}
% \end{minipage}

\begin{figure}
\centering
\includegraphics[width=0.6\textwidth]{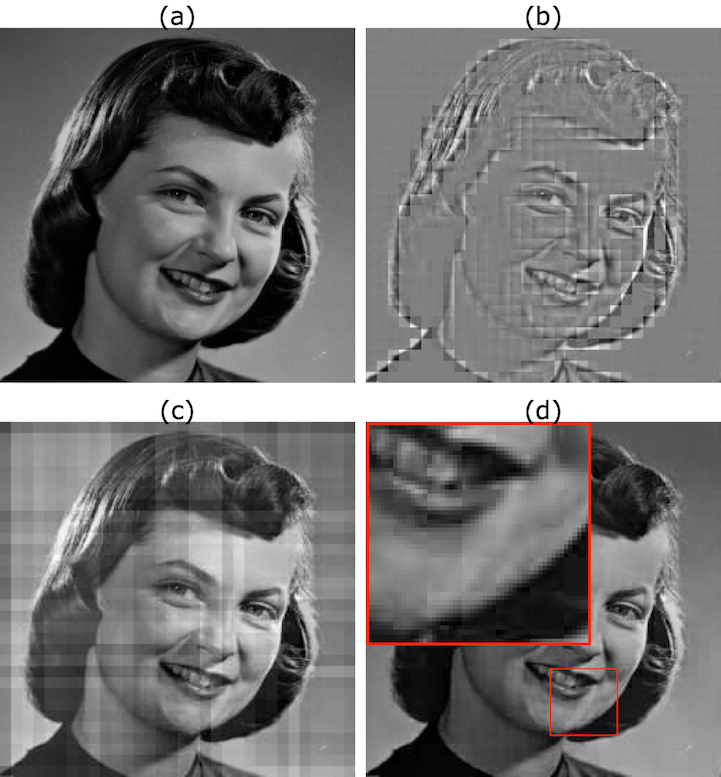}
\caption{Results of the improved DC prediction method proposed in our work: (a) initial JPEG image; (b) JPEG compressed image with no DC coefficients; (c) Recovered image with DC predicted based on methods in~\cite{qiu2019dc}; (d) Recovered image with DC predicted based on method in our work.}
\label{gf4}
\end{figure}

The results of this design are shown in~\figurename~\ref{gf4}. 
There are still some block artifacts that existed if we look into details of the image content (e.g. see the~\figurename~\ref{gf4} (d)) although we could observe a clear improvement compared with the previous method in~\cite{qiu2019dc}. 
The PSNR of the recovery result in~\figurename~\ref{gf4} (d) compared with the JPEG image is around 25 dB. The improvement in visual effects is very obvious but there are still prediction errors leading to the block artifacts.

The limitation of such a method is that there are around 10\% of the $8\times 8$ blocks that contain at least one non-smooth pixel passing with the neighbor blocks. 
In other words, the fundamental reason for such error in DC prediction is that the basic theory as shown in~\cite{uehara2006recovering} we relied on cannot fit for all blocks in the real-world image cases. 
This fact leads to the limitation that no matter how we try to use the average prediction methods to make up the errors of DC coefficients, there are always some blocks that do not fit the basic theory. 
Therefore, the next question is that if we can summarize a statistical model to correct the errors in the DC recovery with a large number of images. 
We answer this question by using the DL models to explore the real-world trends of the pixel passing for correcting the errors in DC prediction. 

\subsection{Analysis and discussion}

Some other related research works have shown the possibility of deploying DL models to solve similar problems in the DCT related questions. 
One of the most popular issues is to use the DL model to enhance the image quality~\cite{ulyanov2018deep} which can also be seen as to remove the noise of the image~\cite{zhang2017learning}. 
However, we are not dealing with the image with the noise defined as the traditional situation since we have the exactly correct AC coefficients compared with the JPEG images, which means we do not have errors in the high-frequency domain compared with the original JPEG images. 
In other words, since we have all correct AC coefficients in the frequency domain and the JPEG image is the target we want to rebuild, there is no noise existed. 
Therefore, directly deploying the image denoising technique is not suitable for our scenario since the denoising techniques are mainly used to remove noises existed in the spatial domain corresponding to the errors of coefficients in the high-frequency bands which do not exist in our case. 

The research direction should be to improve the traditional pixel passing theory based method as the first step and then to use the DL model to solve the block artifacts which is limited by the theory of the traditional approaches. 
The DL model should be used to learn the compensate values between the original JPEG images and the images generated by the DC recovery methods. 
Thus, by using these two steps together as shown in~\figurename~\ref{system1}, these two steps can recover the JPEG images with very limited information at the receiver's end. 
Therefore, we do not use the denoising DL models but deploy the deep residual learning model to realize our target that is to remove the block artifacts generated by the state of the art DC recovery method for JPEG images.

\section{Deep learning model description}
\label{DeepLea}

In this section, we propose a deep CNN model, \textit{Block Artifact Removing Convolutional Neural Networks to Enhance JPEG compression (BARCNN)}, to further enhance the quality of JPEG images. 
Firstly, the objective function will be formulated. 
Secondly, the architecture of our model will be introduced. 
Finally, the implementation will be explained.

\subsection{Formulation}

BARCNN is a deep CNN model used to improve the image quality of the recovered image of the preprocess step in~\figurename~\ref{system1}.  
The purpose is to learn mapping functions between the input image $I$ recovered by the method in Section~\ref{previous} and the original image $J$. 
For achieving this purpose, we improved the DnCNN model in~\cite{zhang2017beyond} by combining the deep residual learning framework~\cite{he2016deep} to design our model. 
The input image of our model can be considered as $I=J+N$, where $J$ is the original JPEG image and $N$ is the image representing block artifacts caused by propagation error mentioned in Section~\ref{background}. 
We aim to learn a residual mapping function $\mathcal{R}(I)$ that satisfies $\mathcal{R}(I)\approx N$. 
Thus, the loss function can be formulated by the averaged \textit{Mean Squared Error (MSE)} as follows:

\begin{equation}
\mathcal{L}(\Theta)=\frac{1}{N}\sum_{i=1}^{N}\Arrowvert \mathcal{R}(I_i;\Theta)-(J_i-I_i) \Arrowvert_{F}^2 
\label{loss1}
\end{equation}

where $\theta$ is the trainable parameters of our model, N is the number of training images. $I_i$ and $J_i$ are the i-th corresponding input image and the original image. 
$\Arrowvert . \Arrowvert_{F}$ denotes the Frobenius norm.
As illustrated in~\figurename~\ref{dnn1}, there two residual blocks in BARCNN model. 
The first block learns a residual mapping $\mathcal{F}(I)=\mathcal{H}(I)-I$. 
The second block learns another residual mapping $\mathcal{G}(\mathcal{H}(I))=\mathcal{K}(I)-\mathcal{H}(I)$. 
Since $\mathcal{K}(I)$ is the approximation of the original image $J$, then the objective residual mapping $\mathcal{R}(I)=\mathcal{G}(\mathcal{H}(I))+\mathcal{F}(I)$. Hence, the loss function in Equation~\ref{loss1} can be further formulated as:

\begin{equation}
\mathcal{L}(\theta)=\frac{1}{N}\sum_{i=1}^{N}\Arrowvert \mathcal{G}(\mathcal{H}(I_i);\Theta)+\mathcal{F}(I_i;\Theta)-(J_i-I_i) \Arrowvert_{F}^2
\label{loss2}
\end{equation}

By using residual learning, our model can learn a residual mapping by a combination of several nonlinear mappings. 
This approach makes it powerful to learn more complex features for image artifacts.

\subsection{Model Architecture}
\label{architecture}
\begin{figure}[!htbp]
\centering
\includegraphics[width=0.4\textwidth]{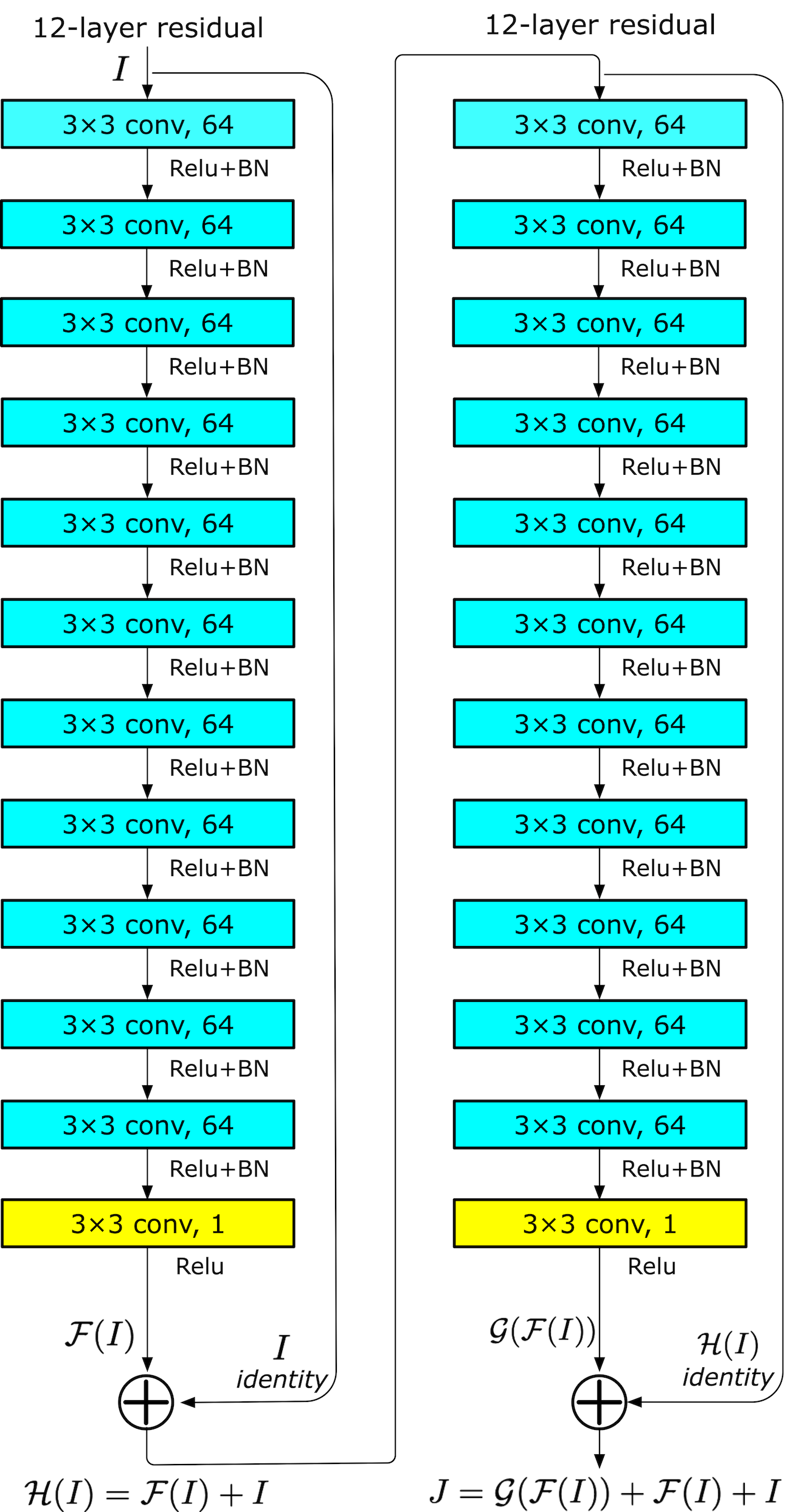}
\caption{The architecture of the proposed BARCNN model, which is inspired by residual learning framework~\cite{he2016deep}. There are two residual blocks used to learn the residual mapping between the input image and noise image that represents block artifacts. Inside each block, there are 12 convolutional layers with ReLU activation. Batch normalization is adopted before activation. More details are shown in Section~\ref{architecture}.}
\label{dnn1}
\end{figure}

Instead of using only one residual mapping in~\cite{zhang2017beyond}, we use two residual blocks to obtain a deeper network. 
As illustrated in~\figurename~\ref{dnn1}, two blocks have identical architecture while each residual block contains 12 convolutional layers and one shortcut connection. 
In a block, all convolutional layers have 64 filters of size $3\times 3$ with a stride of 1, except from the last one that has only one filter of size $3\times 3$ filters with a stride of one, which makes the output have the same shape as the input. 
All layers use rectified linear units (ReLU) as the activation function to produce nonlinearity. 
The batch normalization is also used from the 2nd layer to the 11th layer. According to~\cite{simonyan2014very}, increasing network depth using an architecture with very small convolutional filters can significantly improve the performance of the model. 
Thus, we use filters of size $3\times 3$ and build a neural network with a depth of 24. 
The important features in our model are the use of residual blocks and batch normalization, which make the network easier to be optimized and obtain promising results. The use of residual learning follows the research direction in~\cite{he2016deep}.

In our case, the noise image $N$ represents block artifacts caused by error propagation in the preprocessed DC recovery. 
Compared with the original image $J$, $N=J-I$ has relatively smaller values. It would be easier to optimize a mapping that fits $N$ rather than $J$. 
Hence, nonlinear functions $\mathcal{F}$ and $\mathcal{G}$ are trained to fit the residual mappings. 
In each residual block, there is one shortcut connection, which represents an identity mapping. 
According to~\cite{he2016deep}, this kind of shortcut connection helps to solve the degradation problem that appears in simply stacked nonlinear layers. 
With this architecture, we obtain a deeper CNN model that has higher complexity while is still easy to train. 
The experiment results show that this architecture is effective for our removing block artifacts task.

The use of batch normalization is inspired by~\cite{ioffe2015batch}. 
The change in the distributions of internal nodes of a deep neural network, so-called "internal covariate shift", is considered to having a negative effect on training efficiency. 
In~\cite{santurkar2018does}, the actual impact of batch normalization is to make the optimization landscape significantly smoother. 
This smoothness can result in more efficient training, which motivates us to use this mechanism in our model. 
BARCNN model is trained by using the mini-batch Adam optimization algorithm. 
Thus normalization can be adopted for each mini-batch during each iteration. 
To stabilize the distribution of inputs to the nonlinear function, the batch normalization is performed before Relu activation in each layer. 

\subsection{Implementation}

To train BARCNN model, a big amount of images are needed. Since our model can be applied to images of different sizes, we consider using many small image patches as training data. 
Thus, the input image $I$ of our model is a crop of the image recovered by the method in Section~\ref{previous}. 
The size of the patch should be well chosen to contain significant patterns. Since the DCT transform is performed block by block with a size of $8\times 8$, there will be block artifacts on the boundaries between different blocks. Moreover, errors in predicting DC values result in an intensity difference. 
To better extract these artifacts, we set the size of the patch as 32 to have 16 DCT blocks in one cropped image. 
The experiment results show that patches that we generated contain enough information for removing artifacts.

The cropped images are then fed into BARCNN model. 
Convolutional layers use filters to extract high-level features in the image. Batch normalization helps to stabilize the distribution of those features. And ReLU activation ensures the non-linearity of each layer.
After 12 layers, the output of the first residual block has the same size as the input, which can be considered as the input image plus the related noise image. 
Then the second residual block is used to further tune the noise image through the same procedure. 
The final output is the enhanced image $J$, which will be compared with the reference cropped image through loss function (equation~\ref{loss2}). 
Adam optimization is used to optimize the loss with an adaptive learning rate. 
More details about settings during the training procedure will be introduced in Section~\ref{training}.

\section{Experimentation and evaluation}
\label{Experiment}

In this section, we first list the key implementation details of the BARCNN model training. 
Then, the evaluation results including the visual effects analysis and the statistical analysis are proposed. 
The compression ratio as proposed in~\figurename~\ref{system1} is also calculated on several famous datasets with different image sizes and properties to prove a high compression ratio can be achieved for various kinds of JPEG images by our model.

\subsection{Data preparation}

The images used in our experiments are selected from the LFW dataset~\cite{LFWTech}. 
In total, 5000 human face images are used as dataset and 4000 out of the 5000 images are used as training dataset and the rest of 1000 as testing dataset. 
We apply JPEG compression to all images and set all DC coefficients to zero except the four DC coefficients in corners. 
Then, we use the preprocessing step shown in Section~\ref{problemDef} to predict the DC coefficients and the images with block artifacts are get. 
These images are used as the training dataset by cropping each image into 225 patches of size $32 \times 32$  with a stride of 16. 
Thus, the training data have 900000 cropped images and the testing dataset has 225000 cropped images. 
The reference data have the same number of images cropped from original images. 
All pixel values in images are normalized to values in $[0, 1]$.

\subsection{Training settings}
\label{training}

We use Keras package~\cite{ketkar2017introduction} with Tensorflow~\cite{abadi2016tensorflow} backend to implement our model. 
Weights in all convolutional layers are initialized by random generated orthogonal matrix. 
The parameters of Adam optimizer are set by $\beta_1=0.9, \beta_2=0.999$. 
The initial learning rate is 0.0001. 
Experiments are done on a platform with Intel(R) Xeon(R) CPU E5-2620v3 @ 2.40GHz CPU and three NVIDIA Tesla K80 GPUs. 
The mini-batch size is 256 for each GPU. 
The loss function converges after about 50 epochs.

\subsection{Visual evaluation}

Firstly, we list the visual effects of the recovered images by comparison with the JPEG images, JPEG images with all DC coefficients are zeros and the recovered images based on the improved method based on~\cite{qiu2019dc}. 
As pointed in Section~\ref{training}, we tested 1000 images in the LFW data set to measure the effectiveness of the proposed BARCNN model. 
In~\figurename~\ref{psnr00}, we list four of the images (image id (a), (b), (c), and (d)) picked from the testing data set. 
As we can observe, for the images with all DC coefficients equal to zeros, there are only some blur edges of the original image but all gray scales are lost. 
For the improved method based on~\cite{qiu2019dc}, the basic grayscale gradient is recovered but there are block artifacts due to the errors of the DC values. 
These errors on the DC values are mainly due to the error propagation in the DC guessing process pointed by~\cite{uehara2006recovering,qiu2019dc} which cannot be overcome by the traditional methods. 
Then, we can also observe an obvious improvement based on the BARCNN model on the details. 
The block artifacts including the sharp edges are removed and the whole image quality is improved.

\subsection{Statistical evaluation}

\begin{figure}[!htbp]
    \centering
    \begin{subfigure}{.5\textwidth}
        \centering
        \includegraphics[width=0.9\textwidth]{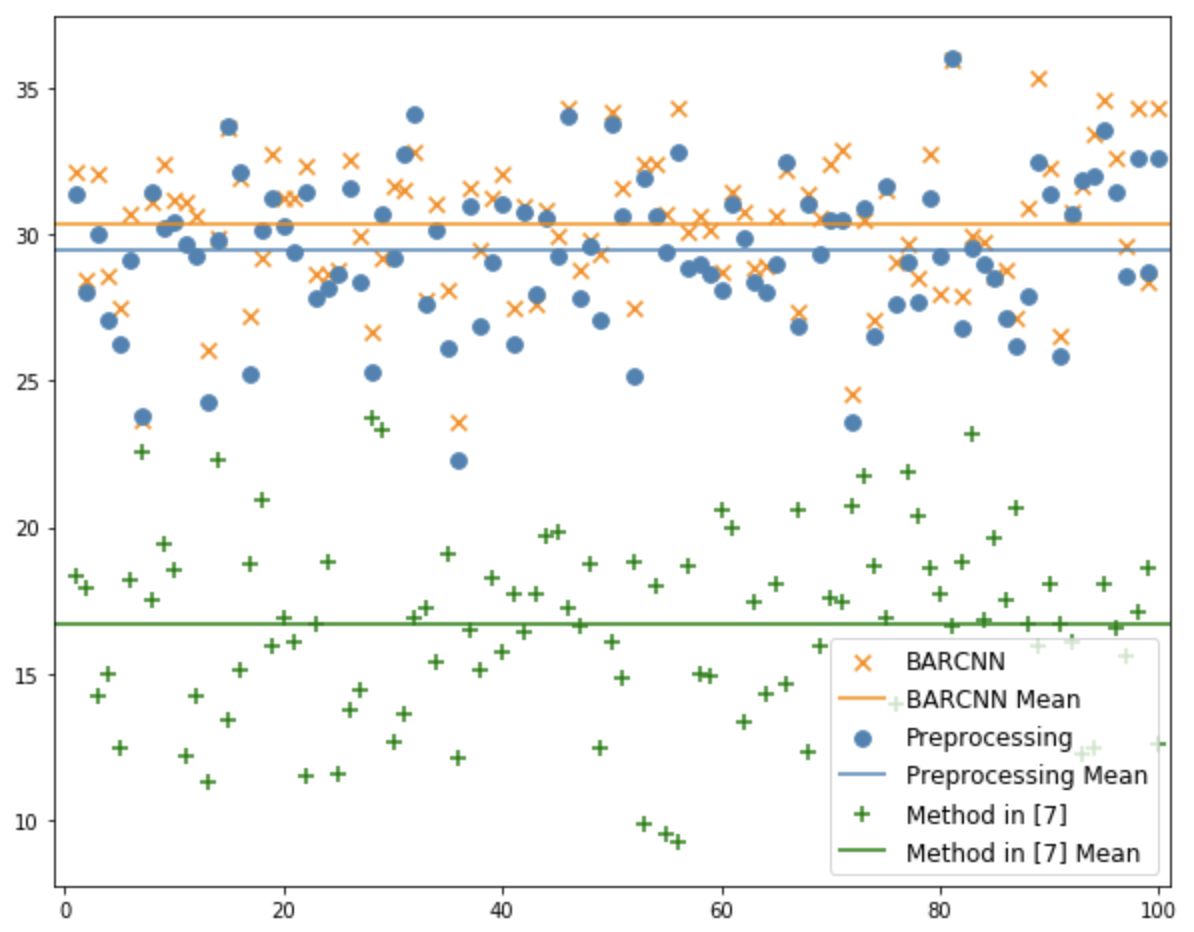}
    \end{subfigure}%
    \begin{subfigure}{.5\textwidth}
        \centering
        \includegraphics[width=0.9\textwidth]{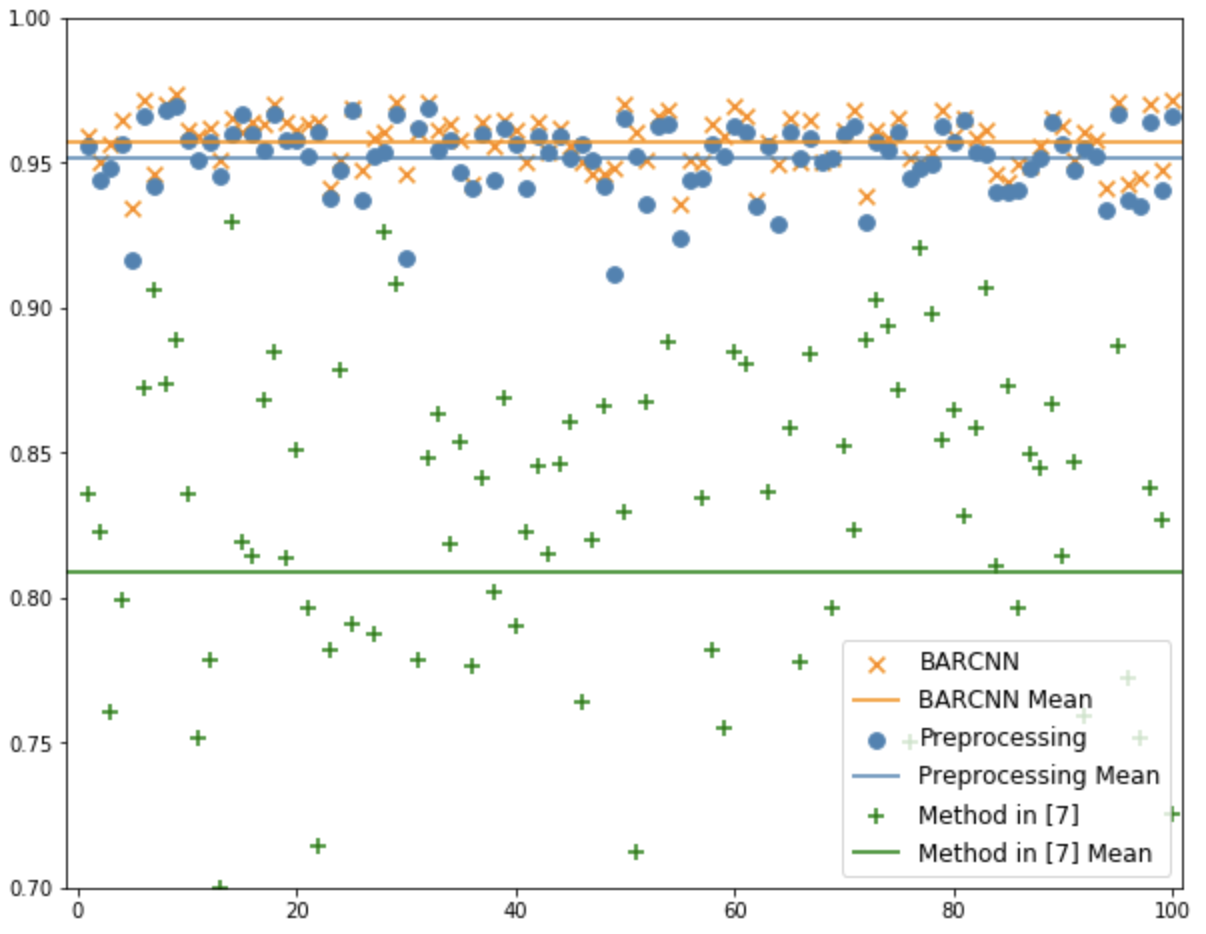}
    \end{subfigure}
    \caption{The statistical results comparison between the recovery methods in~\cite{qiu2019dc}, recovered images of preprocess step, and the BARCNN model on the preprocess results: (a) PSNR; (b) SSIM.}
\label{psnr00}
\end{figure}

\begin{figure*}
\centering
\includegraphics[width=0.96\textwidth]{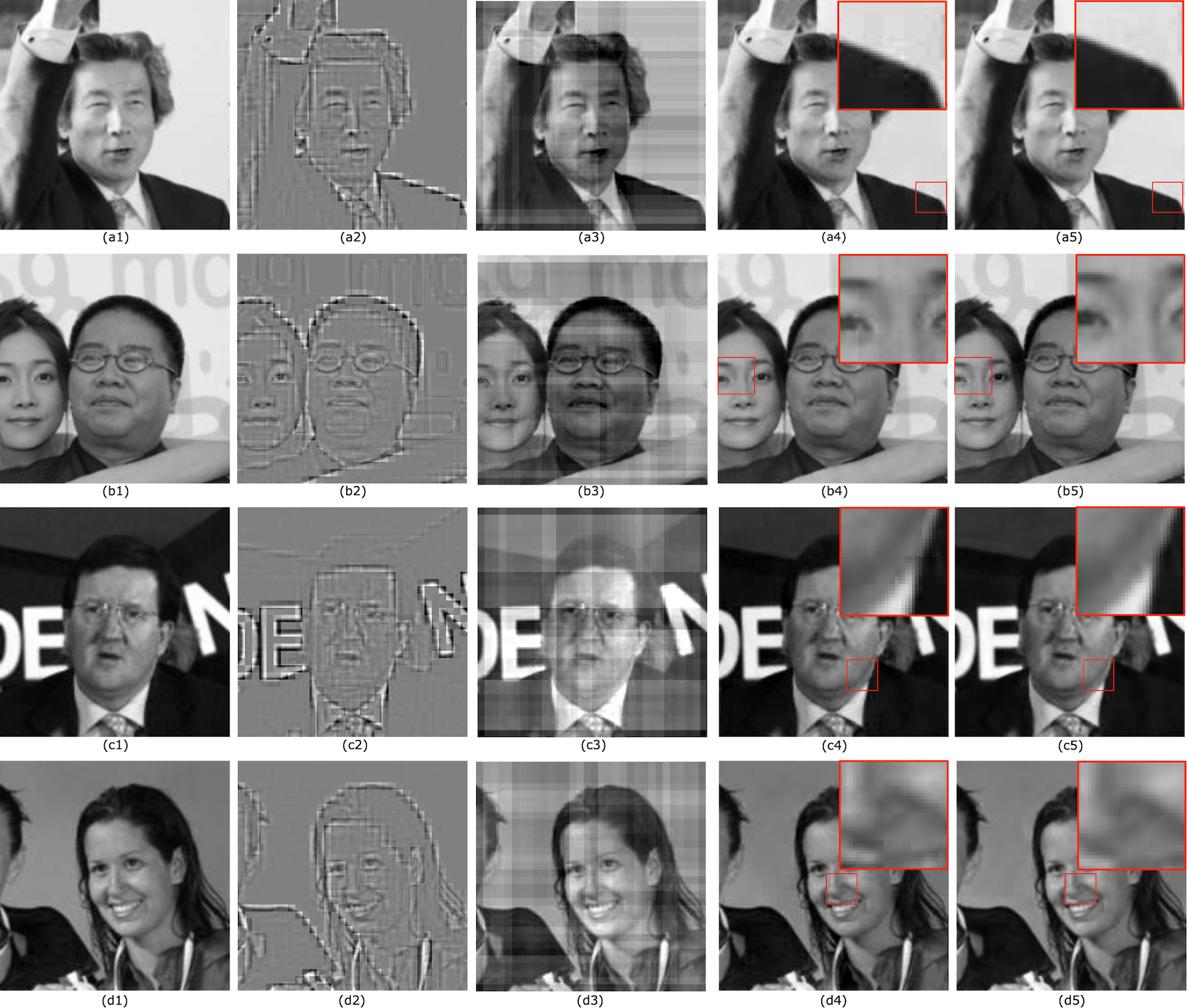}
\caption{The comparison of four example images. \textbf{For the 1st example image}: (a1) original JPEG image; (a2) JPEG compressed image with no DC (PSNR: 10.12 dB, SSIM: 0.5285, compression: 63.58\%); (a3) recovered image by method in~\cite{qiu2019dc} (PSNR: 16.00 dB, SSIM: 82.27\%); (a4) recovered image by preprocess step (PSNR: 31.45 dB, SSIM: 96.69\%); (a5) further corrected image by BARCNN model (PSNR: 34.49 dB, SSIM: 96.98\%). 
\textbf{For the 2nd example image:} (b1) original JPEG image; (b2) JPEG compressed image with no DC (PSNR: 11.55 dB, SSIM: 0.5907, compression: 62.76\%); (b3) recovered image by method in~\cite{qiu2019dc} (PSNR: 16.07 dB, SSIM: 75.53\%); (b4) recovered image by preprocess step (PSNR: 27.12 dB, SSIM: 95.30\%); (b5) further corrected image by BARCNN model (PSNR: 30.22 dB, SSIM: 95.93\%). 
\textbf{For the 3rd example image:} (c1) original JPEG image; (c2) JPEG compressed image with no DC (PSNR: 9.44 dB, SSIM: 0.3957, compression: 63.13\%); (c3) recovered image by method in~\cite{qiu2019dc} (PSNR: 16.03 dB, SSIM: 60.86\%); (c4) recovered image by preprocess step (PSNR: 29.67 dB, SSIM: 94.03\%); (c5) further corrected image by BARCNN model (PSNR: 33.41 dB, SSIM: 95.92\%). 
\textbf{For the 4th example image:} (d1) original JPEG image; (d2) JPEG compressed image with no DC (PSNR: 12.59 dB, SSIM: 0.5285, compression: 63.58\%); (d3) recovered image by method in~\cite{qiu2019dc} (PSNR: 17.91 dB, SSIM: 79.78\%); (d4) recovered image by preprocess step (PSNR: 29.44 dB, SSIM: 95.92\%); (d5) further corrected image by BARCNN model (PSNR: 32.63 dB, SSIM: 96.52\%).
}
\label{evaluation1}
\end{figure*}

We use \textit{Peak Signal-to-Noise Ratio (PSNR)} and \textit{Structural SIMilarity (SSIM)}~\cite{hore2010image} to measure the JPEG image rebuilding results. 
Note that PSNR is normally used to measure the effect of the noise compared with the ground truth images, here the PSNR is measured compared with the JPEG image for two images recovered in two steps as shown in~\figurename~\ref{system1}. 
Firstly we improved the method in~\cite{qiu2019dc} to get a preprocess recovery image as shown in~\figurename~\ref{evaluation1} which contains block artifacts. 
Then the proposed BARCNN model is used to further recover the preprocessed images to test the improvement on the PSNR value. 
In~\figurename~\ref{psnr00} (a), we list the PSNR values of 100 randomly picked images in the test data set (LFW data set) and plot the PSNR values of the preprocess recovered images and the recovered images based on BARCNN model. 
The average PSNR values are also calculated respectively and we can observe an obvious improvement of the PSNR values with the average PSNR values are 29.2 dB and 30.3 dB, respectively.

SSIM is also used to test the image quality after transmission and recovery. 
For the preprocess recovered images, the SSIM on average can reach 0.95 which is the best result in the traditional block-based approach. 
However, the average SSIM of the DL based method is 0.96. 
Moreover, for the preprocess recovered images, the SSIM could be relatively lower until 0.91 while the DL based method is always more than 0.93.

\subsection{Compression ratio evaluation}
\label{compress1}

The baseline scenario is that the sender will directly send the JPEG images with a standard JPEG compression procedure such that the receiver will get a JPEG image. 
The second scenario, as shown in~\figurename~\ref{system1}, uses the same procedure to generate a JPEG image with replacing all DC coefficients as zeros except from the four DC coefficients in the four corner blocks.

\begin{table}[!htbp]
\caption{Compression ratio tests on five datasets.}
\label{compress}
\centering
\begin{tabular}{|c|c|c|c|c|c|}
\hline
    & LFW100   & BSDS100 & General100 & Urban100 & Manga109 \\ \hline
min & 54.93\%  & 44.19\% & 42.59\%    & 46.33\%  & 49.54\%  \\ \hline
max & 70.37\%  & 74.25\% & 78.05\%    & 82.82\%  & 81.13\%  \\ \hline
ave & 63.84\%  & 63.94\% & 63.59\%    & 66.50\%  & 66.38\%  \\ \hline
\end{tabular}
\end{table}

For the proposed scenario, we test the compression ratio on five benchmark dataset. 
The compression ratio is calculated by the ratio between JPEG images with only four DC coefficients and original JPEG images. 
The results are shown in Table~\ref{compress}.
LFW100 contains 100 images selected from our testing dataset. 
Four others are datasets used in~\cite{lai2017deep}. 
Some images have a compression ratio of less than 50\%, which means that they can save half of the storage of JPEG images. The average compression ratio is around 65\%. 
The high compression ratio shows the effectiveness of the proposed JPEG transmission procedure. 
Noticed that in practice, there are several factors that influence the compression ratio, image size, average intensity, spectral distribution, etc.

\section{Discussion}
\label{discuss1}

We are currently exploring the possibility of enhancing the JPEG image compression by recovering the DC coefficients at the receiver's end. 
The main research motivation is to answer the question that if the DC coefficients can be accurately recovered if the AC coefficients are exactly kept proposed in~\cite{uehara2006recovering}. 
The initial research is based on improving the existing approaches but the limitation is very obvious since the basic theory is not 100\% true in real-world image cases. 
Although the PSNR measurement is acceptable the block artifacts are very obvious which became an obstacle that cannot be overcome by the traditional methods. 
Then, we re-defined this obstacle as a question of building a model to fit the pixel distribution property of real-world images which inspired us to deploy the deep learning model. 

However, as pointed in Section~\ref{problemDef}, in our scenario, the block artifacts we want to remove is not the same with the noise of image based on the traditional definition. 
The research proposed in this chapter is to combine the improved methods for accurately DC coefficients recovery, which is the state of the art method, with the deep residual learning model. 
Therefore, by recovering DC coefficients with acceptable accuracy, the deep residual learning model can be used to tinny tune the pixel values to achieve a block-free image with higher PSNR. 
Since the average PSNR compared with the JPEG is more than 30 dB and transmission data ratio is only around 60\% of the initial JPEG image, we believe this research has achieved our initial target that could highly enhance the JPEG compression.

\section{Conclusion}
\label{conclusion}

In this chapter, we proposed an enhanced JPEG compression method by recovering the JPEG image from only four DC coefficients and all AC coefficients at the receiver's end. 
We firstly proposed a state of the art accurate DC recovery method as the preprocess step to generate an image with only the AC coefficients and four DC coefficients. 
Then, in order to get rid of the problem that the observed theory cannot fit all real-world image property, we proposed a deep residual learning model to further remove the block artifacts based on the results of the preprocessed images. 
Our experimentation showed that transmitting only four DC coefficients and all AC coefficients will take only around 60\% data of the original JPEG image while the recovery method can generate an image with average PSNR more than 30 dB for over 1000 JPEG images. 
Therefore, by combining the traditional method and the deep learning method, we presented a practical enhanced JPEG compression method.

\chapter{Wavelet Transformation based Adversarial Example Defenses}
\label{chap:wave}

Recent research works demonstrated that deep neural networks are vulnerable to adversarial examples, which are normally maliciously created by carefully adding deliberate and imperceptible perturbations. 
However, state of the art signal processing-based approaches are either decreasing accuracy on clean datasets or threatened by some specific adversarial examples. 
In this chapter, we propose a novel framework by enhancing not only the input data samples but also the classifiers which are effective against many kinds of adversarial attacks. 
First, we propose a wavelet extension method to extend training data sample dimensions by extracting the image structures and basic elements to be concatenated with the initial image. 
Then, we further add wavelet denoising for the inference step to further reduce the influence of the adaptive adversarial attacks in the white-box scenario. 
By providing intensive experiments on 8 famous adversarial attack methods under all possible scenarios, our proposed method can outperform the two state of the art wavelet-based defense methods.

\section{Introduction}

Deep Learning (DL) based classification methods have been well developed and widely deployed on many real-world systems~\cite{lecun2015deep}. 
However, the robustness of the DL-based classification models is challenged by the existence of Adversarial Examples (AEs)~\cite{yuan2019adversarial}. 
The AEs are generated by adding carefully designed perturbations which are usually imperceptible compared with clean data samples but can mislead the classifiers with surprising results~\cite{goodfellow2014explaining}. 
Specifically, the initial AE generated is an image with very small perturbations and very limited imperceptible to human eyes but can decrease the accuracy of the neural network classifiers to almost zero. 

In recent years, many research works on various kinds of AE generation methods are proposed. 
Initially, AEs are generated by calculating model gradients such as Fast Gradient Sign Method (FGSM)~\cite{goodfellow2014explaining} is calculated based on the sign of the gradient of the classification loss with respect to the input sample. 
Later, other approaches are proposed to enhance the AE generation methods by not only calculating gradients but also using optimization algorithms~\cite{carlini2017towards} which enhanced the generation of AEs. 
On the other hand, recent works showed that AEs threats classifiers for not only the image classification tasks, but also other tasks such as Natural Language Processing (NLP)~\cite{jia2017adversarial} or malware detection~\cite{grosse2017adversarial}. 
Thus, the existence of AEs significantly threatens the reliability and robustness of the classifiers. 

The countermeasures dealing with AEs are also developed based on different adversary's knowledge including white-box, black-box, and gray-box scenarios~\cite{meng2017magnet}. 
The white-box scenario of DL service is that all details of the model are known including training data, model architectures, hyperparameters, numbers of layers, activation functions, and model weights. 
The black-box scenario also exists as online ML services like Google Cloud AI~\cite{yuan2019adversarial} with only the input data sample and output (label or confidence score) are known. 
The gray-box scenario is also proposed and similar to the white-box scenario except that the parameters keep unknown. 
Initially, AE generation methods are based on calculating gradients of the classification loss with respect to parameters. 
However, they can be also used to attack the black-box scenario due to the transferability of AEs~\cite{papernot2016transferability} which makes the defense schemes in black-box scenario vulnerable. 

One approach to designing the defense scheme is to perform differentiable transformations on an input image before classification. 
However, AEs can still be generated by taking the gradient of a class probability with respect to input pixels through both the model and the transformation. 
More complicated methods following this idea were proposed like~\cite{prakash2018deflecting,shaham2018defending} to use non-differentiable transformations but was proved still vulnerable by~\cite{athalye2018robustness}. 

In this chapter, we propose a novel approach, WD-WavExt, to enhance the robustness of neural networks by not only enhancing the data sample but also the model. 
First, we propose a Wavelet Extension (WavExt) method to extract the image structures and basic elements to combine with the initial image as one tensor for the training that can enhance the robustness against tiny perturbations. 
Then, for the inference step, we propose the Wavelet-based Denoising (WD) as a preprocessing step which is different from the training preprocess step to further reduce the influence ofAEs. 

The roadmap of this chapter is as follows. 
In Section 2, we briefly introduce the background information including the current strategies of defending AEs. 
In Section 3, we present WD-WavExt with details.
In Section 4, we evaluate WD-WavExt comparing with two states of the art approaches under different scenarios. 
We conclude in Section 6.

\section{Research Background}

In this section, we first point out the necessity of choosing defense against AE by recovering its ground truth labels rather than the only detection on if the input is an AE or not. 
We also briefly introduce the current defense strategies on AEs including the backgrounds, defense preliminaries, and current main approaches.

\subsection{Problem of Countermeasures: Detect or Defend?}

The very initial purpose of generating AE is to mislead the classifiers usually by adding imperceptible perturbations. 
Therefore, the basic ideas of dealing with AEs can be concluded as either to detect one AE and reject it or to classify the AE as its correct label. 

For instance, the detection of AE can be seen as a function that decides whether the input data sample is an AE. 
In~\cite{metzen2017detecting}, the authors proposed to train a classifier to distinguish between normal input data samples and AEs. 
The limitation of such approaches, as pointed in~\cite{meng2017magnet}, is obvious since it will require the defender to understand the attacker such as acquiring future AEs or knowing the process for generating AEs. 
This approach seems not to be able to provide a generalized defend scheme for AEs since if the detector was trained with slightly perturbed AEs, the detector had high false-positive rates because it decided many normal examples as adversarial. 

On the other hand, however, considering a non-AE input data sample, the classifier may also misclassify its label since the accuracy of classifiers is usually not 100\% which is the same effect as the AE generation.  
Therefore, it is questionable to design methods to detect whether an input is an AE or not since detection AE does not equal to increase the model's robustness. 
In~\cite{huang2019model}, the authors claimed that their method can effectively detect the AEs. 
First, in the practical scenario, the inputs are more complicated since they can be normal input, normal input with wrong prediction labels, AEs, non-AE inputs with noise, etc. 
Papers like~\cite{huang2019model} only evaluated the accuracy of detecting AEs but how to recover the correct label is still challenging since a classifier with high accuracy on agnostic inputs including AEs is more important. 
Such detection approaches are not practically useful for maintaining the performance of classifiers.

Therefore, from a viewpoint based on a classification-oriented purpose, we claim the valuable point should be to get the correct label of the data samples processed with AE generation methods.

\subsection{Defense Strategies on Adversarial Attacks}

%What is the state of art Adversarial Examples defense
The current designs of countermeasures on AEs can be summarized as following two main types of ideas including reactive and proactive. 
The proactive approaches aim to make DNN models more robust against the AE generations. 
This approach can be represented as the augmentation on model $F$ as $F'$ such that $F'(\widetilde{x}) = F(x)$ such as network distillation~\cite{papernot2016distillation}. 
However, $F'$ can be considered as the novel target function and AEs can be further generated according to the novel $F'$~\cite{carlini2017towards}. 
On the other hand, the reactive approaches aim to detect or prevent the AEs after the DNN models are built. 
This approach can be presented as: given a classification model $F$ and an image $\widetilde{x}$, which may either be an original image $x$, or an adversarial image $\widehat{x}$, the goal of such reactive approach is to design a transformation function $\tau$ such that $F(\tau(\widetilde{x})) = F(x)$. 

One of the reactive approaches is to design a transformation function $\tau$ such that $F(\tau(\widetilde{x})) = F(x)$. 
Since AEs are normally tiny but fragile perturbations which can be influenced by signal processing techniques such as wavelet transformation shown in~\cite{prakash2018deflecting} or denoising networks~\cite{liao2018defense}. 

The main motivation of such transformation-based defenses is to influence the malicious perturbations with one more layer of perturbations. 
By performing either noise removing or image quality enhancement operations~\cite{mustafa2019image}, the limitation of such approaches is obvious. 
On the one hand, such transformation-based defenses are that with very tiny defensive perturbations, the influence of AEs cannot be significantly removed. 
On the other hand, intensive defensive perturbations will lead to an obvious decrease in the accuracy of clean images that disobey the initial defend purpose. 

This problem is because the visual element in one image is made of visual structures, basic visual elements, and visual details. 
Such transformation-based defenses are always suffering from the issue that the trade-off between removing AEs and controlling the changes made on initial visual structures, basic visual elements, and visual details.

\section{Proposed method}

In this section, we first briefly introduce our initial motivation for designing the AE defense method. 
Then, the theory and practical details of the wavelet extension by decomposition and reconstruction are discussed. 
After we pick the certain wavelet filter, we build the AE defense method and all design details are presented. 

\subsection{Basic Designs}

For an image $x$, we note its AE as $\widetilde{x}=x+\delta$ with $\delta$ the adversarial perturbation. 
Let F denotes the function of target classifier, the problem of adversarial attacks can be formulated as the following:

\begin{equation}
    min  \lVert\delta\rVert ,\: s.t. \: F(\widetilde{x})\neq F(x)
\end{equation}

Attackers try to generate $\widetilde{x}$ misclassified by target classifier while keeping $\widetilde{x}$ visually as imperceptible as possible.

\begin{equation}
    \max_{\tau} P(F(\tau(\widetilde{x}))=F(\tau(x)))
    \label{eq:transform}
\end{equation}

\begin{equation}
    \max_{F'} P(F'(\widetilde{x})=F'(x))
    \label{eq:model}
\end{equation}

The defenses against AEs are mainly based on two ideas. 
The first idea is to apply transformations on input images to reduce the influence of carefully added perturbations, as in eq.~\ref{eq:transform}.
The transformations can be non-differentiable and non-invertible, which makes adversaries difficult to get the gradients of the target model through back-propagation.
The second idea is to modify the architecture of the target model to increase robustness, as in eq.~\ref{eq:model}. 
The modified model keeps high accuracy with clean inputs while it is effective on certain AEs.

Our idea is to combine these two schemes to benefit from their strengths.
We also introduce a new mechanism to increase the difficulty of generating AEs by considering the following problem:
\begin{align}
    \max_{F',\tau_1, \tau_2} &\ P(F'(\tau_1(\widetilde{x}), \tau_2(\widetilde{x}))=F'(\tau_1(x), \tau_2(x)))
\end{align}

In this case, the target classifier $F'$ takes two parts into consideration at the same time. 
In practice, the adversary can only provide $\widetilde{x}$ as input to the entire system.
Then different transformations $\tau_1, \tau_2$ are applied to $\widetilde{x}$ to extract features of different levels. 
This mechanism requires adversaries to generate AEs that are effective on both parts. 
The problem is now to find the proper transformations that satisfy this situation.
Based on this principle, we introduce our defense method in the following subsections.

\subsection{WavExt: Wavelet Extension}

The basic idea of the wavelet transform is to represent any arbitrary function $f$ as a superposition of wavelets~\cite{antonini1992image}. 
Discrete Wavelet Transformation (DWT) decomposes $f$ into different scale levels, where each level is then further decomposed with a resolution adapted to the level. 
One way to achieve such a decomposition writes $f$ as an integral over $a$ and $b$ of
$\psi^{a,b}$ with appropriate weighting coefficients where $\psi$ are mother waves that satisfies the property of $\int \psi(x)dx = 0$. 

In practice, one prefers to write $f$ as a discrete superposition (sum rather than integral): $f = \sum c^{a,b} \psi^{a,b}$. 
Therefore, with the proper choice of the wavelet filter, the image can be decomposed into several scales representing different elements such as visual content structures, details, etc which are usually used for compression ~\cite{skodras2001jpeg}. 

In our work, we propose Wavelet Extension (WavExt) to deploy the wavelet decomposition and reconstruction to extract and rebuild the image elements and structures to assist the classification. 
As we all know, the existing AE generation methods mainly try to add tiny perturbation on images to mislead the classifiers which modify little on the basic content structures. 
Therefore, we could use the wavelet decomposition and reconstruction to extend the space of the RGB image from $N\times N\times3$ into $N\times N\times6$ by adding the reconstructed image structures.

%make supplementary data for training to improve the robustness of classifiers with a certain wavelet. 
%method to 

%(a wavelet-based extension, WavExt, with size $N\times N\times3$ added after the image)

Particularly, we have tested several known wavelet transform methods and pick 'db4' as the filter since the 'db4' filter can best represent the image structure and basic elements according to~\cite{mallat1999wavelet}. 
After one level of two dimensional DWT, there are four sub-bands generated with each size of $floor(\frac{N-1}{2})+n$. 
This value in the CIFAR-10 dataset~\cite{krizhevsky2009learning} is 19 since $N = 32$ of image size and $n = 4$ as the filter length ('db4'). 
We cut the LL band of each RGB layer from $19\times 19$ to $16\times 16$ then perform the Bicubic interpolation to resize the extracted low frequency back to $32 \times 32$. 
Three color layers are all performed in such a way and the result, a $32 \times 32 \times 3$ tensor will be added after the input image as the actual input tensor for the training. 
The wavelet extension algorithm is shown in Algorithm 1 and the preprocess on training data samples is given in~\figurename~\ref{fig:modeltrain} (a). 

\begin{figure}
\centering
\includegraphics[width=0.6\textwidth]{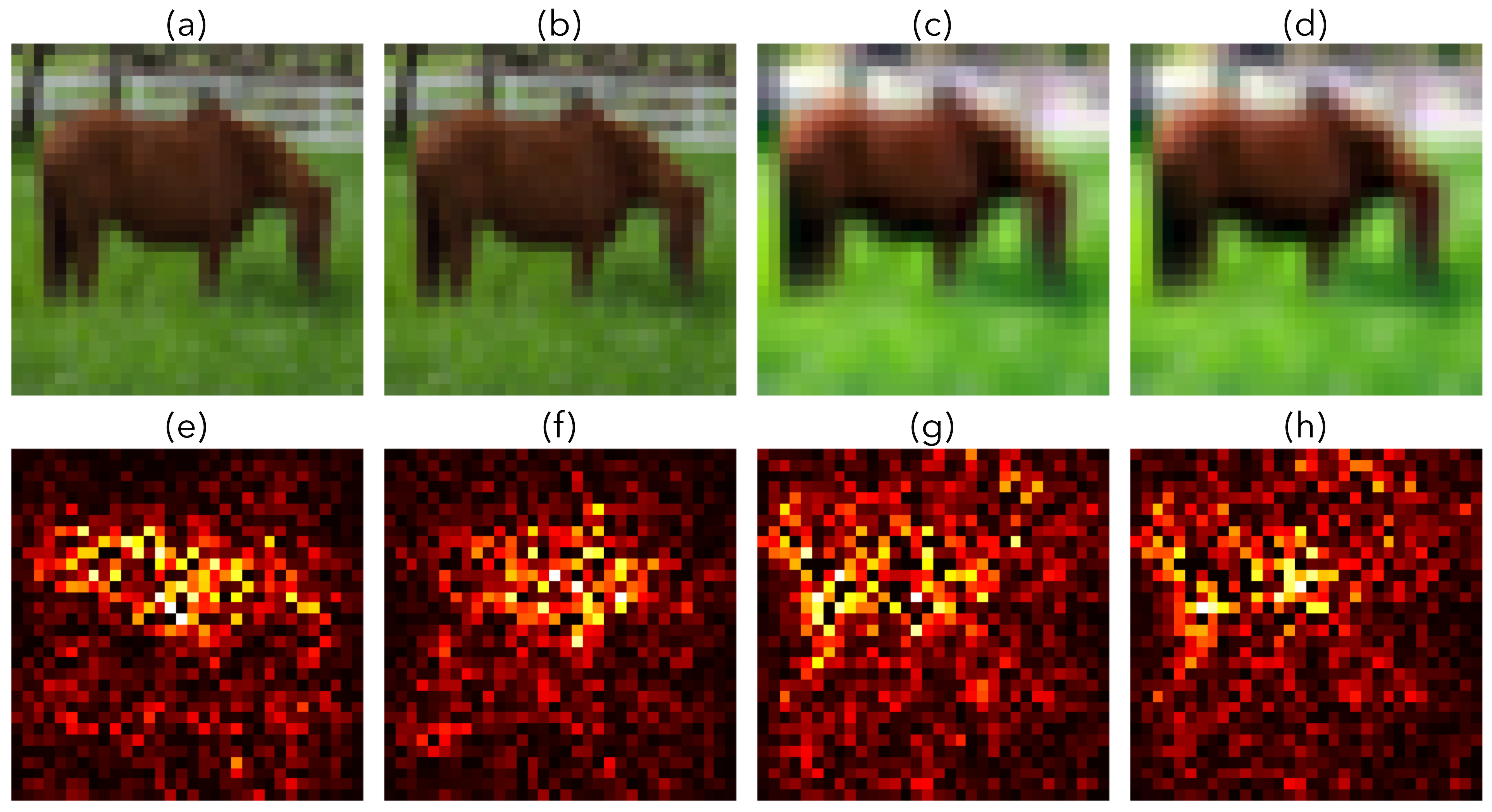}
\caption{An example to show \textbf {similar saliency maps between (g) and (h)}. (a) and (e): clean image and its saliency map; (b) and (f): corresponding AE image and its saliency map; (c) and (g): Wavelet extension of image (a) and its saliency map; (d) and (h): Wavelet extension of AE image (b) and its saliency map.}
\label{fig:wavelet1}
\end{figure}

In order to prove the effectiveness of using wavelet extension, we calculate the saliency map~\cite{tramer2017space}. 
As shown in~\figurename~\ref{fig:wavelet1} (e) and (f), as the saliency maps for the clean image(\figurename~\ref{fig:wavelet1} (a)) and corresponding AE (\figurename~\ref{fig:wavelet1} (b)), have obvious difference that causes misclassification. 
However, the saliency maps of their extracted structures with wavelet decomposition and reconstruction (\figurename~\ref{fig:wavelet1} (g) and (h), respectively) are relatively very similar.

\begin{algorithm}[tb]
\caption{Process of WavExt Defense.}
\label{alg:algorithm}
\textbf{Input}: Image $x$, size $N\times N\times 3$\\
\textbf{Output}: Class $\mathcal{C}$
\begin{algorithmic}[1] %[1] enables line numbers
\STATE LL, LH, HL, HH = wavelet-decomposition(I)
\STATE LL is cropped to the size $N/2\times N/2 \times 3$
\STATE The value of LL is normalised to [0, 255]
\STATE $x_{Ext}$ = Bicubic-Interpolation(LL)
\IF {Inference}
\STATE $x_{Extended}$ = concat(WD($x$), $x_{Ext})$
\ELSE 
\STATE $x_{Extended}$ = concat($x$, $x_{Ext}$)
\ENDIF
\STATE $C = \arg\max~Classifier(x_{Extended})$
\end{algorithmic}
\end{algorithm}

\subsection{WD-WavExt: Wavelet Denoising and Wavelet Extension}

%In this section, we propose a Wavelet-based Extension (WavExt) to extend the actual training data samples by concatenating the initial $N \times N \times 3$ image with 

After the wavelet decomposition and reconstruction in Section 3.2, our DNN model will be trained with an input size of $N \times N \times 6$ although the input of the system is still with the size of $N \times N \times 3$. 
Therefore, considering the initial motivation of using wavelet extension, the extended 3 layer tensor is the extracted image structures and basic elements used for assisting the classification. 
The extended 3 layer tensor in our work is proved to increase the accuracy of the classification which adds the robustness of the DNN models. 

Considering the practical threat model, once the attackers have the knowledge of our model details, one attempt might be to generate an AE from our DNN model and get an AE tensor with size $N \times N \times 6$. 
However, in our design, the input must be an image with the size of $N \times N \times 3$ so the attacker may try to use the first three-layer of the AE tensor with size $N \times N \times 6$ to mislead the classification. 
After we receive this AE with size $N \times N \times 3$, we will perform the wavelet decomposition and reconstruction and concatenate the generated $N \times N \times 3$ tensor with the AE as the input. 

\begin{figure}
\centering
\includegraphics[width=0.9\textwidth]{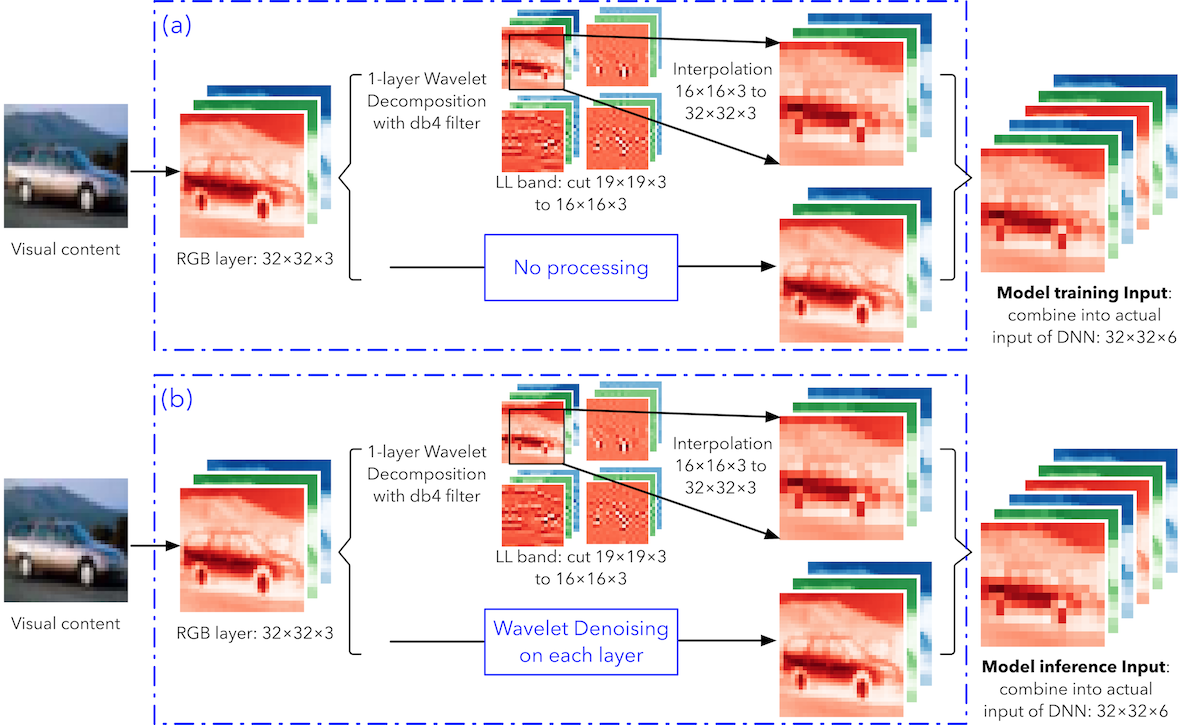}
\caption{Proposed method with different preprocessing steps during training and inference.}
\label{fig:modeltrain}
\end{figure}

Furthermore, we actually use different preprocessing step to inference input image with size $N \times N \times 3$ as shown in~\figurename~\ref{fig:modeltrain} (b). 
The motivation of using different preprocessing steps for training and inference is to introduce the non-differentiable process at the inference step to increase the difficulty for gradient-based AE generation methods. 
The point here is gradient-based AE generation methods will calculate the gradients according to the trained model parameters which will not have the best-optimized result since the inference step has a different preprocessing step.

In this work, we use the wavelet-based denoising method named BayesShrink as shown in~\cite{chang2000adaptive}. 
For an image $X$ with $N$ pixels, this is given by $\sigma \sqrt{2\textrm{log}N}$, where $\sigma$ is the variance of the noise to be removed and is a hyper-parameter. 
In our work, we model the threshold for each wavelet coefficient as a Generalized Gaussian Distribution (GGD). 
The optimal threshold is then assumed to be the value which minimizes the expected mean square error as follows. 

\begin{equation}
    T_{h} * (\sigma_{x}, \beta ) = \underset{T_{h}}{\textup{\textrm{argmin}}} \: E (\widehat{X} -X)^2 \approx \frac{\sigma^2}{\sigma_x}
\end{equation}

where $\sigma_{x}$ and $\beta$ are parameters of the GGD for each wavelet sub-band and the approximation can be then calculated as $\frac{\sigma^2}{\sigma_x}$. 
Within a certain range of $\beta$, BayesShrink could effectively remove artificial noise while preserving the perceptual features of natural images~\cite{prakash2018deflecting}. 

In summary, for the training steps, the enhance method is only relying on WavExt while for the inference steps, the enhanced method is relying on WavExt combined with the Wavelet Denoising (WD-WavExt). 
Of course, we admit that the BayesShrink based denoising step may be further replaced with other better methods within our proposed system. 

\begin{table*}[!htbp]
\centering
\caption{The Top-1 accuracy in the presence of AEs generated by various adversarial attacks on undefended model.}
\begin{tabular}{cccclcc}
\hline
\textbf{Attack}   & $L_{\inf}$ & $L_2$ & \textbf{Undefended} & \textbf{WA}   & \textbf{PD} & \textbf{WD-WavExt} \\ \hline
Clean             & 0.0           & 0.0         & 1.0                 & 0.89          & 0.92          & \textbf{0.97}      \\
FGSM-1 ($\epsilon=0.03$)     & 0.030          & 1.65        & 0.18                & 0.65          & 0.27          & \textbf{0.67}      \\ 
FGSM-2 ($\epsilon=0.005$)    & 0.005         & 0.28        & 0.39                & 0.84          & 0.84          & \textbf{0.94}      \\ 
I-FGSM-1 ($\epsilon=0.03$)   & 0.030          & 0.88        & 0.0                 & \textbf{0.68} & 0.22          & 0.55               \\ 
I-FGSM-2 ($\epsilon=0.005$)  & 0.005         & 0.21        & 0.21                & 0.85          & 0.86          & \textbf{0.93}      \\ 
MI-FGSM-1 ($\epsilon=0.03$)  & 0.030          & 1.28        & 0.0                 & \textbf{0.56} & 0.06          & 0.51               \\ 
MI-FGSM-2 ($\epsilon=0.005$) & 0.005         & 0.25        & 0.29                & 0.84          & 0.86          & \textbf{0.92}      \\ 
JSMA              & 0.832          & 4.12        & 0.0                 & 0.60          & 0.49          & \textbf{0.68}      \\ 
DeepFool          & 0.015         & 0.12        & 0.0                 & 0.87          & 0.95          & \textbf{0.95}      \\ 
LBFGS             & 0.018         & 0.15        & 0.0                 & 0.86          & 0.92          & \textbf{0.97}      \\ 
CW                & 0.011         & 0.09        & 0.0                 & 0.86          & 0.95          & \textbf{0.97}      \\ 
BPDA              & 0.015         & 0.72        & 0.0                 & 0.83 & 0.81           & \textbf{0.86}               \\ \hline
\end{tabular}
\label{table:eval1}
\end{table*}

\section{Evaluations}

In this section, we evaluate the robustness of our defense against various adversarial attacks.
We first introduce our experiment settings about the target model and dataset. 
Then we evaluate our defense in two different scenarios, AEs generated by adversarial attacks on the undefended model, and, AEs generated by adaptive adversarial attacks under the knowledge of our defense method.

\subsection{Training Settings}

In our experiments, we consider the image classification task on the CIFAR-10 dataset. 
There are 50000 images for training and 10000 images for testing. 
Each image is of size $32\times 32\times 3$ and belongs to one of ten classes.
All pixel values are normalized to be in $[0, 1]$.
The target classifier is the ResNet-29~\cite{he2016identity}.
It consists of in total of 29 layers containing three bottleneck residual blocks with channel size 64, 128, 256, respectively.
The model is trained to reach the Top-1 accuracy of $92.27\%$ under the evaluation of all test images.

\subsection{Evaluation Settings}

As pointed in~\figurename~\ref{fig:modeltrain}, the model with our defense takes a $N\times N\times 3$ image of three RGB layers as input and applies wavelet extension to extend this image into a tensor of size $N\times N\times 6$ to be fed into the classifier. 

\begin{figure}
\centering
\includegraphics[width=0.6\textwidth]{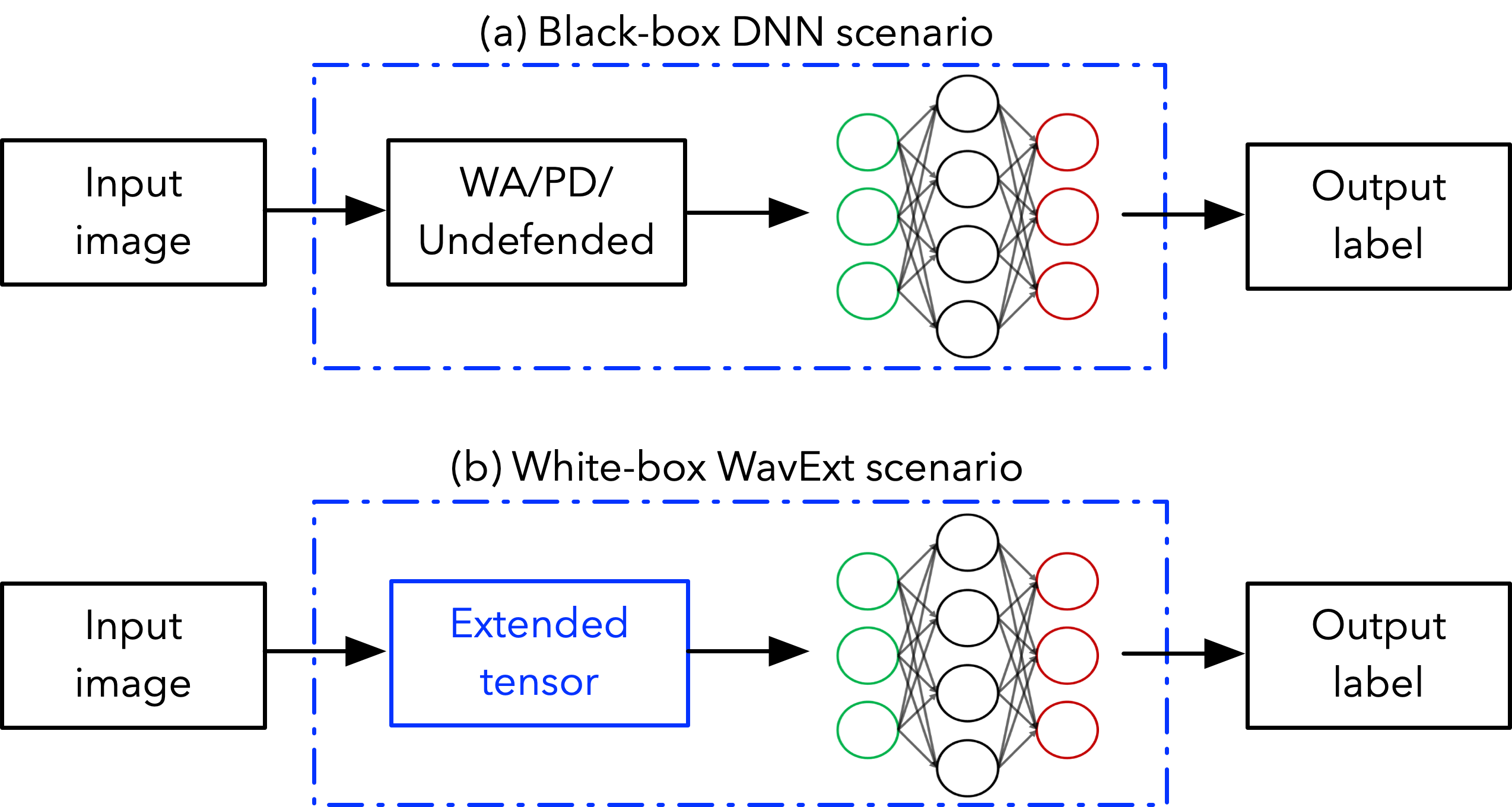}
\caption{Evaluation scenarios: (a) black-box scenario; (b) white-box scenario.}
\label{fig:evaluation1}
\end{figure}

In order to evaluate our method along with the model, we test according to the following configurations. 
Since the input of WavExt defended classifier is changed to size $N\times N\times 6$, we train another ResNet-29 model and the Top-1 accuracy on test images reaches $91.96\%$ (DNN model shown in~\figurename~\ref{fig:evaluation1} (a)). 
We compare our method with two states of the art wavelet transform-based defense approaches, Wavelet Approximation (WA)~\cite{shaham2018defending} and Pixel Deflection (PD)~\cite{prakash2018deflecting}. 

The comparison is made up of two aspects including the black-box scenario and the white-box scenario (n~\figurename~\ref{fig:evaluation1}). 
The first black-box scenario is that the attacker is aware of only the input data sample and the output label in~\figurename~\ref{fig:evaluation1} (a) which fits the definition for the black-box scenario according to~\cite{meng2017magnet}. 
Of course, due to the transferability of the neural networks, the AE attacks can still perform on this DNN model. 
We use 8 most famous AE generation methods (including FGSM, I-FGSM, MI-FGSM~\cite{dong2018boosting}, L-BFGS, JSMA, DeepFool, CW and BPDA~\cite{athalye2018obfuscated}) to generate AEs to test the classification accuracy on the undefended DNN model, DNN model with WA protection, DNN model with PD protection, and our proposed WD-WavExt method (results in Section 4.3). 

Second, we also evaluate the white-box scenario (results in Section 4.4) that the attackers have all the knowledge of our model's training details as shown in~\figurename~\ref{fig:modeltrain} (a) including all the model's parameters. 
Since the proposed WavExt uses different preprocessing and extending methods on the training data and on the inference data, the definition of white-box in~\cite{meng2017magnet} means the inference steps are unknown to attackers. 
However, still, we evaluate the situation that even the attackers know the inference steps which is the extreme situation and the test results are given in Section 4.5. 

%The defended model is trained under the same settings as the target model and the Top-1 accuracy on test images reaches $91.96\%$. 
Moreover, in order to focus on the performance in the presence of AEs, the images that are misclassified by the undefended model before applying attacks are ignored. 
Since some of the attacks are computationally intensive, we randomly selected 100 test images that are correctly predicted by the original model as the evaluation set.

\subsection{Evaluation in Black-box Scenario}

%To evaluate the robustness of our method against different attacks, we consider 8 different types of adversarial attacks including FGSM, I-FGSM, MI-FGSM~\cite{dong2018boosting}, L-BFGS, JSMA, DeepFool, CW, and BPDA~\cite{athalye2018obfuscated}. 
In the black-box scenario, the details of performing AE attacks are given as follows. 
For FGSM, I-FGSM and MI-FGSM, we generate AEs with different scales of distortion $\epsilon=0.005, 003$ under the $L_{inf}$ constraints. 
For the rest of AE generation methods, we generate AEs until the accuracy of the undefended model drops to 0.
For targeted attacks, we randomly generated the target that is different from the original one. 
All AEs are generated targeting the undefended target classifier.
To clearly show the magnitude of AEs distortion compared with the original images, we calculate the average normalized $L_{\inf}$ and $L_2$ distance. 

We also re-experiment the two wavelet transformation-based approaches following the details given as follows. 
The WA uses level-1 wavelet approximation on the input image to get low-resolution representation. 
The PD randomly selects pixels and replaces them by other pixels randomly selected within a small window followed by wavelet denoising to reduce adversarial noise. 
Since PD is a stochastic process, we perform 10 runs and take the majority of prediction to evaluate the accuracy as proposed in the original paper. 
We adjust the window size and the number of deflections.

The Top-1 accuracy in the presence of AEs generated on the undefended model is shown in~\tableautorefname~\ref{table:eval1}. 
Some attacks like DeepFool, L-BFGS, and CW make the accuracy of the undefended model drops to 0 with little $L_2$ distortion. 
Our defense is very effective towards all 8 kinds of AE attacks, which outperforms both WA and PD in most cases.  
According to the experimentation, for attacks with larger $L_2$ distortion, like I-FGSM-1, and MI-FGSM-1, PD defense is no longer effective. 
However, our defense still shows resistance in this case because of the wavelet-based denoising and the wavelet-based extension representation for the image structures and basic elements. 

\begin{table*}[!htbp]
\centering
\caption{The Top-1 accuracy in the presence of AEs generated by adapting adversarial attacks to our defense.}
\begin{tabular}{ccccc}
\hline
\textbf{Attack}   & $L_{\inf}$ & $L_2$ &  \textbf{WavExt} & \textbf{WD-WavExt} \\ \hline
FGSM-1    & 0.030         & 1.65        & 0.18            & \textbf{0.35}      \\ 
FGSM-2    & 0.005         & 0.28        & 0.48            & \textbf{0.62}      \\ 
I-FGSM-1   & 0.030         & 0.81        & 0.00            & \textbf{0.04}               \\ 
I-FGSM-2  & 0.005         & 0.21        & 0.24            & \textbf{0.48}      \\ 
MI-FGSM-1 & 0.030         & 1.25        & 0.02            & \textbf{0.06}               \\ 
MI-FGSM-2 & 0.005         & 0.24        & 0.28            & \textbf{0.50}      \\ 
JSMA              & 0.898         & 4.95        & 0.19            & \textbf{0.68}      \\ 
DeepFool          & 0.015         & 0.12        & 0.85            & \textbf{0.95}      \\ 
LBFGS             & 0.017         & 0.15        & 0.77            & \textbf{0.97}      \\ 
CW                & 0.012         & 0.09        & 0.87            & \textbf{0.97}      \\ \hline
\end{tabular}
\label{table:eval2}
\end{table*}

\subsection{Evaluation in White-box Scenario}

According to the definition of the white-box scenario in~\cite{meng2017magnet}, the attackers know all the details of the target classifier even including the parameters. 
Then, the AEs generation based on these methods will produce AEs with a size of $N \times N\times 6$ which fits our actual DNN model. 
However, since we force to accept the input data sample with a size of $N \times N\times 3$ so we define the threat model in this scenario is that the attackers will take an RGB image and use the WavExt to get a $N \times N\times 6$ tensor with the first 3 layers remains the same as visual content but the rest 3 layers are wavelet extension representing the image structures. 
Then, attackers will generate a $N \times N\times 6$ tensor accordingly as the AEs but we only accept the input data sample of $N \times N\times 3$. 
In this case, it is pointless to try to perform the reverse calculation on the lossy wavelet extension (last 3 layers) back since the reverse calculation in~\figurename~\ref{fig:modeltrain} (a) is impossible with a cut operation. 
So we assume the attacker will pick the first 3 layers of the $N \times N\times 6$ AE tensor as the actual input to mislead our model. 

Since we have a different size of input with the actual input for the classifier, the existing approaches like WA or PD cannot fit our case which cannot be used for comparison. 
We generate AEs by 7 methods and adapt the AEs into an input data sample of $N \times N\times 3$ by separating the first three layers. 
We compare the WavExt method which is using the AEs with their wavelet extension and the WD-WavExt which is to combine the denoising results with the wavelet extension as shown in~\tableautorefname~\ref{table:eval2}. 
We can observe that, after adaptive attacks, our defense still shows resistance against some attacks, especially for DeepFool, LBFGS, and CW. 
%We also show the benefit by applying different processes during the training phase and inference phase. 
Therefore, we proved that (as shown in Section 3.3) by using the wavelet denoising to actual input during the inference phase, the robustness is increased and maintained. 

\subsection{Further Discussion with BPDA}

Although most previous researches didn't consider the case that adversaries are fully aware of the defense method, we claim that it is important to evaluate the robustness in this circumstance. 
In this section, we assume the attackers understand everything including the details of BayesShrink wavelet denoising used in the inference step. 
Therefore, the classification process can still be summarized as $y=f(x)$ which $x$ is the input data sample with size of $N \times N\times 3$ and $f$ is a mapping function. 

In this case, many previous gradient-based attacks such as FGSM have problems handling the non-differentiable function $f$ which includes the wavelet denoising and extension. 
In other words, calculating one AE with size $N \times N\times 3$ directly from calculating the gradient will not work in this case which is called obfuscated gradient~\cite{athalye2018obfuscated}. 

We do notice that there is one possibility, Backward Pass Differentiable Approximation (BPDA)~\cite{athalye2018robustness}, as a powerful attack that may potentially overcome this difficulty brought by defense methods. 
The idea of BPDA is to calculate gradients by replacing the non-differentiable transformation with the identity function on the back-propagation process. 
By iteratively adapting the defense method and observing the change of prediction, BPDA modifies the original image to generate AEs.

\begin{figure}
\centering
\includegraphics[width=0.6\textwidth]{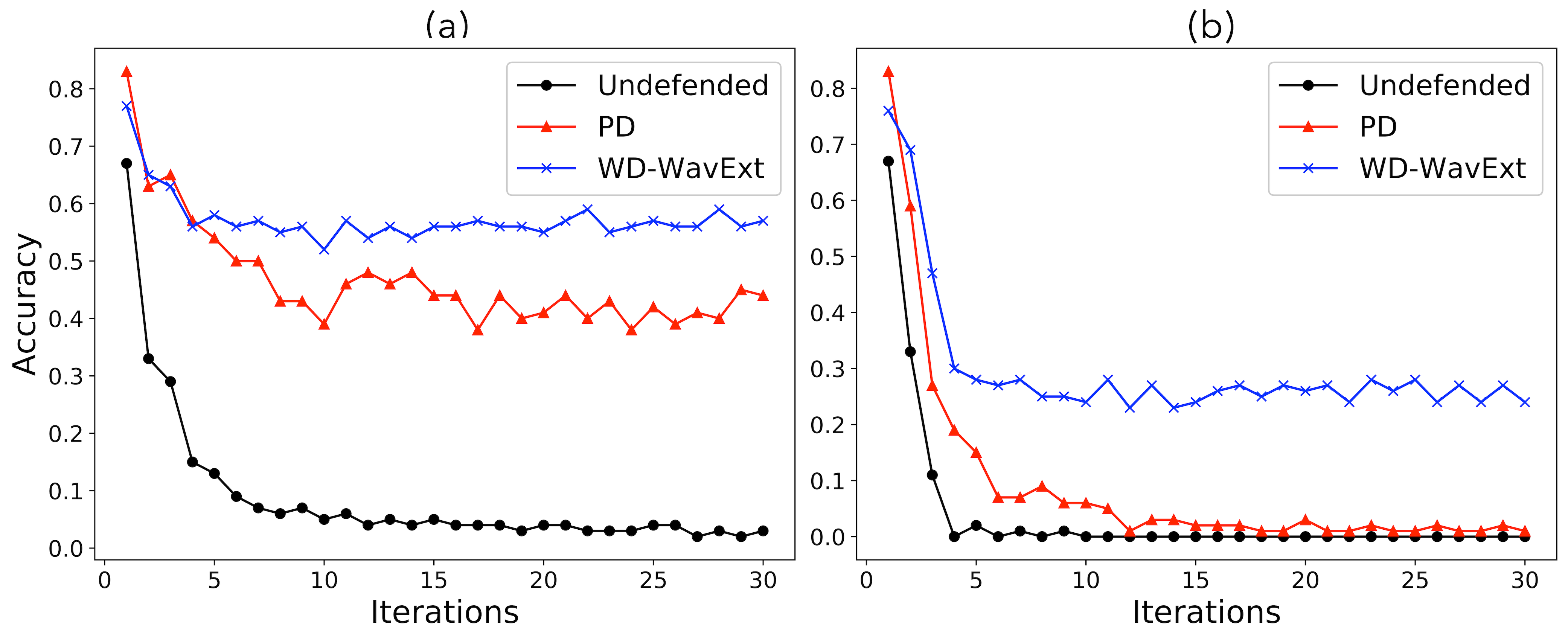}
\caption{Evaluation of PD and WD-WavExt against the BPDA attack. (a) Under the constraint of $L_{\inf}=0.008$ (b) Under the constraint of $L_{\inf}=0.016$.}
\label{fig:bpda}
\end{figure}

In~\cite{athalye2018robustness}, the author claims that they use BPDA to break PD defense and the accuracy of PD can be significantly decreased. 
We also apply BPDA to attack our defense under the constraints of two magnitudes as shown in~\tableautorefname~\ref{fig:bpda}. 
The results show that our defense is more robust compared with PD under the same constraints.
With enough iterations and larger constraints, BPDA can still make the accuracy of the WD-WavExt defended model drop dramatically. 
However, this result shows that the architecture of Wd-WavExt can still increase the difficulty for BPDA to generate AEs and outperforms the state of the art PD method. 

In summary, we are not trying to propose a method that can generally defend all kinds of AEs. 
However, according to our intensive experimentation under all possible scenarios, we believe our framework of building models and preprocess data samples provides a novel direction to increase the DNN robustness against many famous AE attacks compared with the state of the art defense methods.  
Particularly, we also provide an idea for defending BPDA by introducing transformations that can't be easily induced during back-propagation.

\section{Conclusion}

In this chapter, we proposed a novel framework for defending the AEs. 
First, we extract the image structures, basic elements as supplementary data to concatenate with images for improving the robustness of target classifiers. 
By defining a novel process, WavExt, we changed the dimension from input data samples to actual classifier inputs which increased difficulty for attackers. 
Also, we use different preprocessing steps for training and inference to defend gradient-based AE generation methods by avoiding direct access to gradients in white-box scenarios. 
We demonstrated the robustness of our defense in both black-box and white-box scenarios with intensive experimentation by outperforming two states of the art approaches. 

\chapter{Future Work and Conclusion}
\label{chap:discuss}

In this report, we proposed two image processing applications that combine the transformations in the frequency domain and deep learning techniques.
These applications show the interest of processing images in the frequency domain. 

Firstly, we proposed a novel method to enhance JPEG compression by transmitting much less data while maintaining good image quality.
This enhanced compression method can considerably increase the efficiency of image data transmission. 
Nowadays, the amount of image data continues to increase due to the development of social networks. 
Transmissions of all these image data become a big challenge to the networks.
By using the proposed enhanced JPEG compression method, we can reduce the amount of data transmitting through the networks.
This only requires some additional computation on the edge devices, e.g. computers, cellphones, tablets, etc.
Fortunately, the devices have already become much powerful than ever before and can easily handle the computation.
There are still some potentials to improve our method.
It is possible to discards more information during transmission and then recover the image at the receiver's end. 
For example, discarding some AC coefficients that represent high-frequency content, i.e. details.
It is also possible to use more advanced deep learning architecture to better recover the image. 

Secondly, we proposed a novel framework for defending the AEs through wavelet extension. 
We demonstrated the robustness of our defense in both black-box and white-box scenarios. 
In practice, there are many deep learning-based applications in image processing, e.g. face recognition, object detection, etc. 
However, the security of these applications is not totally ensured due to the existence of adversarial examples. 
Our method can be used to enhance the robustness of current applications against potential adversary attacks. 
It can be easily adapted to every kind of neural networks by only changing parameters in the first layer and adding a preprocessing.
We also notice that our method can be improved in several aspects.
1) The wavelet transform can be used in a different way. 
Currently, we assume that the adversarial perturbations appear mostly in the high-frequency domain, which might not be exact.
If we can further explore the distribution of adversarial perturbations in the frequency domain, we can reduce their impact by using wavelet decomposition differently.
2) In order to defend the BPDA attack, more work needs to be done. 
We already demonstrated that our method is more robust than other defense under the BPDA attack.
However, it still shows limited robustness. 
It is possible to introduce more transformations at the same time to shrink the effective space of adversarial examples.
This will make it difficult for adversaries to generate adversarial examples, at least with small perturbations.

All in all, we show the potential of using transformations including discrete cosine transform and wavelet transform together with deep learning models on image processing applications. 
Although the end-to-end deep learning framework has its advantages and simplicity, we show that the use of frequency domain can be beneficial in some cases. 
This provides a novel direction when we design the architecture of deep learning models.

\bibliographystyle{unsrt}
\bibliography{template}

\begin{thebibliography}{10}

\bibitem{manjunath1996texture}
Bangalore~S Manjunath and Wei-Ying Ma.
\newblock Texture features for browsing and retrieval of image data.
\newblock {\em IEEE Transactions on pattern analysis and machine intelligence},
  18(8):837--842, 1996.

\bibitem{devore1992image}
Ronald~A DeVore, Bj{\"o}rn Jawerth, and Bradley~J Lucier.
\newblock Image compression through wavelet transform coding.
\newblock {\em IEEE Transactions on information theory}, 38(2):719--746, 1992.

\bibitem{krizhevsky2012imagenet}
Alex Krizhevsky, Ilya Sutskever, and Geoffrey~E Hinton.
\newblock Imagenet classification with deep convolutional neural networks.
\newblock In {\em Advances in neural information processing systems}, pages
  1097--1105, 2012.

\bibitem{zhang2017beyond}
Kai Zhang, Wangmeng Zuo, Yunjin Chen, Deyu Meng, and Lei Zhang.
\newblock Beyond a gaussian denoiser: Residual learning of deep cnn for image
  denoising.
\newblock {\em IEEE Transactions on Image Processing}, 26(7):3142--3155, 2017.

\bibitem{balle2016end}
Johannes Ball{\'e}, Valero Laparra, and Eero~P Simoncelli.
\newblock End-to-end optimized image compression.
\newblock {\em arXiv preprint arXiv:1611.01704}, 2016.

\bibitem{qiu2019user}
Han Qiu, Hassan Noura, Meikang Qiu, Zhong Ming, and Gerard Memmi.
\newblock A user-centric data protection method for cloud storage based on
  invertible dwt.
\newblock {\em IEEE Transactions on Cloud Computing}, 2019.

\bibitem{miao2016performance}
Wang Miao, Geyong Min, Yulei Wu, Haozhe Wang, and Jia Hu.
\newblock Performance modelling and analysis of software-defined networking
  under bursty multimedia traffic.
\newblock {\em ACM Transactions on Multimedia Computing, Communications, and
  Applications (TOMM)}, 12(5s):77, 2016.

\bibitem{ahmed1974discrete}
Nasir Ahmed, T~Natarajan, and Kamisetty~R Rao.
\newblock Discrete cosine transform.
\newblock {\em IEEE transactions on Computers}, 100(1):90--93, 1974.

\bibitem{schwarz2007overview}
Heiko Schwarz, Detlev Marpe, and Thomas Wiegand.
\newblock Overview of the scalable video coding extension of the {H}. 264/{AVC}
  standard.
\newblock {\em IEEE Transactions on circuits and systems for video technology},
  17(9):1103--1120, 2007.

\bibitem{zeng2013tutorial}
Jin Zeng, Oscar~C Au, Wei Dai, Yue Kong, Luheng Jia, and Wenjing Zhu.
\newblock A tutorial on image/video coding standards.
\newblock In {\em IEEE Signal and Information Processing Association Annual
  Summit and Conference (APSIPA)}, pages 1--7, 2013.

\bibitem{wallace1992jpeg}
Gregory Wallace.
\newblock The {JPEG} still picture compression standard.
\newblock {\em IEEE transactions on consumer electronics}, 38(1):xviii--xxxiv,
  1992.

\bibitem{chen2017dc}
Chen Chen, Zexiang Miao, Xiandong Meng, Shuyuan Zhu, and Bing Zeng.
\newblock {DC} coefficient estimation of intra-predicted residuals in {HEVC}.
\newblock {\em IEEE Transactions on Circuits and Systems for Video Technology},
  28(8):1906--1919, 2017.

\bibitem{qiu2019dc}
Han Qiu, Gerard Memmi, Xuan Chen, and Jian Xiong.
\newblock {DC} coefficient recovery for {JPEG} images in ubiquitous
  communication systems.
\newblock {\em Future Generation Computer Systems}, 96:23--31, 2019.

\bibitem{uehara2006recovering}
Takeyuki Uehara, Reihaneh Safavi-Naini, and Philip Ogunbona.
\newblock Recovering {DC} coefficients in block-based {DCT}.
\newblock {\em IEEE Transactions on Image Processing}, 15(11):3592--3596, 2006.

\bibitem{kresch1999fast}
Renato Kresch and Neri Merhav.
\newblock Fast {DCT} domain filtering using the {DCT} and the {DST}.
\newblock {\em IEEE transactions on Image Processing}, 8(6):821--833, 1999.

\bibitem{qiu2019dc2}
Han Qiu, Qinkai Zheng, Meikang Qiu, and Gerard Memmi.
\newblock Dc coefficients recovery from ac coefficients in the jpeg compression
  scenario.
\newblock In {\em International Conference on Smart Computing and
  Communication}, pages 266--276. Springer, 2019.

\bibitem{li2011recovering}
Shujun Li, Andreas Karrenbauer, Dietmar Saupe, and C-C~Jay Kuo.
\newblock Recovering missing coefficients in {DCT}-transformed images.
\newblock In {\em 18th IEEE International Conference on Image Processing
  (ICIP)}, pages 1537--1540, 2011.

\bibitem{qiu2018ssic}
Han Qiu, Nathalie Enfrin, and Gerard Memmi.
\newblock A case study for practical issues of {DCT} based bitmap selective
  encryption methods.
\newblock In {\em IEEE International Conference on Security of Smart Cities,
  Industrial Control System and Communications}, 2018.

\bibitem{pennebaker1992jpeg}
William~B Pennebaker and Joan~L Mitchell.
\newblock {\em JPEG: Still image data compression standard}.
\newblock Springer Science \& Business Media, 1992.

\bibitem{ulyanov2018deep}
Dmitry Ulyanov, Andrea Vedaldi, and Victor Lempitsky.
\newblock Deep image prior.
\newblock In {\em Proceedings of the IEEE Conference on Computer Vision and
  Pattern Recognition}, pages 9446--9454, 2018.

\bibitem{zhang2017learning}
Kai Zhang, Wangmeng Zuo, Shuhang Gu, and Lei Zhang.
\newblock Learning deep {CNN} denoiser prior for image restoration.
\newblock In {\em Proceedings of the IEEE conference on computer vision and
  pattern recognition}, pages 3929--3938, 2017.

\bibitem{he2016deep}
Kaiming He, Xiangyu Zhang, Shaoqing Ren, and Jian Sun.
\newblock Deep residual learning for image recognition.
\newblock In {\em Proceedings of the IEEE conference on computer vision and
  pattern recognition}, pages 770--778, 2016.

\bibitem{simonyan2014very}
Karen Simonyan and Andrew Zisserman.
\newblock Very deep convolutional networks for large-scale image recognition.
\newblock {\em arXiv preprint arXiv:1409.1556}, 2014.

\bibitem{ioffe2015batch}
Sergey Ioffe and Christian Szegedy.
\newblock Batch normalization: Accelerating deep network training by reducing
  internal covariate shift.
\newblock {\em arXiv preprint arXiv:1502.03167}, 2015.

\bibitem{santurkar2018does}
Shibani Santurkar, Dimitris Tsipras, Andrew Ilyas, and Aleksander Madry.
\newblock How does batch normalization help optimization?
\newblock In {\em Advances in Neural Information Processing Systems}, pages
  2483--2493, 2018.

\bibitem{LFWTech}
Gary~B. Huang, Manu Ramesh, Tamara Berg, and Erik Learned-Miller.
\newblock Labeled faces in the wild: A database for studying face recognition
  in unconstrained environments.
\newblock Technical Report 07-49, University of Massachusetts, Amherst, October
  2007.

\bibitem{ketkar2017introduction}
Nikhil Ketkar.
\newblock Introduction to keras.
\newblock In {\em Deep Learning with Python}, pages 97--111. Springer, 2017.

\bibitem{abadi2016tensorflow}
Mart{\'\i}n Abadi, Paul Barham, Jianmin Chen, Zhifeng Chen, Andy Davis, Jeffrey
  Dean, Matthieu Devin, Sanjay Ghemawat, Geoffrey Irving, Michael Isard, et~al.
\newblock Tensorflow: A system for large-scale machine learning.
\newblock In {\em 12th {USENIX} Symposium on Operating Systems Design and
  Implementation ({OSDI} 16)}, pages 265--283, 2016.

\bibitem{hore2010image}
Alain Hore and Djemel Ziou.
\newblock Image quality metrics: {PSNR} vs. {SSIM}.
\newblock In {\em IEEE International Conference on Pattern Recognition}, pages
  2366--2369, 2010.

\bibitem{lai2017deep}
Wei-Sheng Lai, Jia-Bin Huang, Narendra Ahuja, and Ming-Hsuan Yang.
\newblock Deep laplacian pyramid networks for fast and accurate
  super-resolution.
\newblock In {\em Proceedings of the IEEE conference on computer vision and
  pattern recognition}, pages 624--632, 2017.

\bibitem{lecun2015deep}
Yann LeCun, Yoshua Bengio, and Geoffrey Hinton.
\newblock Deep learning.
\newblock {\em nature}, 521(7553):436, 2015.

\bibitem{yuan2019adversarial}
Xiaoyong Yuan, Pan He, Qile Zhu, and Xiaolin Li.
\newblock Adversarial examples: Attacks and defenses for deep learning.
\newblock {\em IEEE transactions on neural networks and learning systems},
  2019.

\bibitem{goodfellow2014explaining}
Ian~J Goodfellow, Jonathon Shlens, and Christian Szegedy.
\newblock Explaining and harnessing adversarial examples.
\newblock {\em arXiv preprint arXiv:1412.6572}, 2014.

\bibitem{carlini2017towards}
Nicholas Carlini and David Wagner.
\newblock Towards evaluating the robustness of neural networks.
\newblock In {\em 2017 IEEE Symposium on Security and Privacy (SP)}, pages
  39--57. IEEE, 2017.

\bibitem{jia2017adversarial}
Robin Jia and Percy Liang.
\newblock Adversarial examples for evaluating reading comprehension systems.
\newblock {\em arXiv preprint arXiv:1707.07328}, 2017.

\bibitem{grosse2017adversarial}
Kathrin Grosse, Nicolas Papernot, Praveen Manoharan, Michael Backes, and
  Patrick McDaniel.
\newblock Adversarial examples for malware detection.
\newblock In {\em European Symposium on Research in Computer Security}, pages
  62--79. Springer, 2017.

\bibitem{meng2017magnet}
Dongyu Meng and Hao Chen.
\newblock Magnet: a two-pronged defense against adversarial examples.
\newblock In {\em Proceedings of the 2017 ACM SIGSAC Conference on Computer and
  Communications Security}, pages 135--147, 2017.

\bibitem{papernot2016transferability}
Nicolas Papernot, Patrick McDaniel, and Ian Goodfellow.
\newblock Transferability in machine learning: from phenomena to black-box
  attacks using adversarial samples.
\newblock {\em arXiv preprint arXiv:1605.07277}, 2016.

\bibitem{prakash2018deflecting}
Aaditya Prakash, Nick Moran, Solomon Garber, Antonella DiLillo, and James
  Storer.
\newblock Deflecting adversarial attacks with pixel deflection.
\newblock In {\em Proceedings of the IEEE conference on computer vision and
  pattern recognition}, pages 8571--8580, 2018.

\bibitem{shaham2018defending}
Uri Shaham, James Garritano, Yutaro Yamada, Ethan Weinberger, Alex Cloninger,
  Xiuyuan Cheng, Kelly Stanton, and Yuval Kluger.
\newblock Defending against adversarial images using basis functions
  transformations.
\newblock {\em arXiv preprint arXiv:1803.10840}, 2018.

\bibitem{athalye2018robustness}
Anish Athalye and Nicholas Carlini.
\newblock On the robustness of the cvpr 2018 white-box adversarial example
  defenses.
\newblock {\em arXiv preprint arXiv:1804.03286}, 2018.

\bibitem{metzen2017detecting}
Jan~Hendrik Metzen, Tim Genewein, Volker Fischer, and Bastian Bischoff.
\newblock On detecting adversarial perturbations.
\newblock {\em arXiv preprint arXiv:1702.04267}, 2017.

\bibitem{huang2019model}
Bo~Huang, Yi~Wang, and Wei Wang.
\newblock Model-agnostic adversarial detection by random perturbations.
\newblock In {\em Proceedings of the 28th International Joint Conference on
  Artificial Intelligence}, pages 4689--4696. AAAI Press, 2019.

\bibitem{papernot2016distillation}
Nicolas Papernot, Patrick McDaniel, Xi~Wu, Somesh Jha, and Ananthram Swami.
\newblock Distillation as a defense to adversarial perturbations against deep
  neural networks.
\newblock In {\em 2016 IEEE Symposium on Security and Privacy (SP)}, pages
  582--597. IEEE, 2016.

\bibitem{liao2018defense}
Fangzhou Liao, Ming Liang, Yinpeng Dong, Tianyu Pang, Xiaolin Hu, and Jun Zhu.
\newblock Defense against adversarial attacks using high-level representation
  guided denoiser.
\newblock In {\em Proceedings of the IEEE Conference on Computer Vision and
  Pattern Recognition}, pages 1778--1787, 2018.

\bibitem{mustafa2019image}
Aamir Mustafa, Salman~H Khan, Munawar Hayat, Jianbing Shen, and Ling Shao.
\newblock Image super-resolution as a defense against adversarial attacks.
\newblock {\em IEEE Transactions on Image Processing}, 29:1711--1724, 2019.

\bibitem{antonini1992image}
Marc Antonini, Michel Barlaud, Pierre Mathieu, and Ingrid Daubechies.
\newblock Image coding using wavelet transform.
\newblock {\em IEEE Transactions on image processing}, 1(2):205--220, 1992.

\bibitem{skodras2001jpeg}
Athanassios Skodras, Charilaos Christopoulos, and Touradj Ebrahimi.
\newblock The jpeg 2000 still image compression standard.
\newblock {\em IEEE Signal processing magazine}, 18(5):36--58, 2001.

\bibitem{mallat1999wavelet}
St{\'e}phane Mallat.
\newblock {\em A wavelet tour of signal processing}.
\newblock Elsevier, 1999.

\bibitem{krizhevsky2009learning}
Alex Krizhevsky, Geoffrey Hinton, et~al.
\newblock Learning multiple layers of features from tiny images.
\newblock Technical report, Citeseer, 2009.

\bibitem{tramer2017space}
Florian Tram{\`e}r, Nicolas Papernot, Ian Goodfellow, Dan Boneh, and Patrick
  McDaniel.
\newblock The space of transferable adversarial examples.
\newblock {\em arXiv preprint arXiv:1704.03453}, 2017.

\bibitem{chang2000adaptive}
S~Grace Chang, Bin Yu, and Martin Vetterli.
\newblock Adaptive wavelet thresholding for image denoising and compression.
\newblock {\em IEEE transactions on image processing}, 9(9):1532--1546, 2000.

\bibitem{he2016identity}
Kaiming He, Xiangyu Zhang, Shaoqing Ren, and Jian Sun.
\newblock Identity mappings in deep residual networks.
\newblock In {\em European conference on computer vision}, pages 630--645.
  Springer, 2016.

\bibitem{dong2018boosting}
Yinpeng Dong, Fangzhou Liao, Tianyu Pang, Hang Su, Jun Zhu, Xiaolin Hu, and
  Jianguo Li.
\newblock Boosting adversarial attacks with momentum.
\newblock In {\em Proceedings of the IEEE conference on computer vision and
  pattern recognition}, pages 9185--9193, 2018.

\bibitem{athalye2018obfuscated}
Anish Athalye, Nicholas Carlini, and David Wagner.
\newblock Obfuscated gradients give a false sense of security: Circumventing
  defenses to adversarial examples.
\newblock {\em arXiv preprint arXiv:1802.00420}, 2018.

\end{thebibliography}


\end{document}